\documentclass[final,pdftex,11pt]{mfart}

\RequirePackage[OT1]{fontenc}
\RequirePackage{amsthm}
\RequirePackage[cmex10]{amsmath}
\RequirePackage{natbib}
\RequirePackage[colorlinks,citecolor=blue,urlcolor=blue]{hyperref}
\RequirePackage{hypernat}

\usepackage{amsmath}
\usepackage{amsfonts}
\usepackage{amssymb}
\usepackage{graphicx}

\startlocaldefs \numberwithin{equation}{section}
\theoremstyle{plain}
\newtheorem{lemma}{Lemma}[section]
\newtheorem{theorem}{Theorem}[section]
\newtheorem{corollary}{Corollary}[section]
\newtheorem{proposition}{Proposition}[section]
\newtheorem{definition}{Definition}[section]
\newtheorem{example}{Example}[section]
\newtheorem{remark}{Remark}[section]
\endlocaldefs

\begin{document}

\begin{frontmatter}
\fontsize{14}{17}
\textbf{MULTIDIMENSIONAL DYNAMIC RISK MEASURE VIA CONDITIONAL $g$-EXPECTATION}

\runtitle{multidimensional risk measure}

\begin{aug}
\author{\snm{Yuhong} \fnms{Xu}
\ead[label=e1]{yuhong.xu@hotmail.com}}

\runauthor{Yuhong Xu}

\address{This work is partially supported by the Project 111 (No. B12023), NSF of China (No. 10921101) and the National Basic
Research Program of China (973 Program) (No. 2007CB814900)
(Financial Risk). The author would like to show many thanks to Prof.
S. Peng for his guidance. Prof. S. Peng is a great master who shows
generosity and constantly reflects encouragement to his students
until they succeed. I am also grateful to the two reviewers. It must
take much time for them to read this paper and give me many useful comments
and suggestions which substantially strengthen the paper.}

\address{ Address correspondence to Yuhong Xu, Institute
of Mathematics, Shandong University, Jinan 250100, P.R.China.
\textbf{E-mail}: \printead{e1}}
\end{aug}
\vspace{2mm}
\textit{Shandong University} \vspace{6mm}

\begin{abstract}
This paper deals with multidimensional dynamic risk measures induced
by conditional $g$-expectations. A notion of multidimensional
$g$-expectation is proposed to provide a multidimensional version of
nonlinear expectations. By a technical result on explicit
expressions for the comparison theorem, uniqueness theorem and
viability on a rectangle of solutions to multidimensional backward
stochastic differential equations, some necessary and sufficient
conditions are given for the constancy, monotonicity, positivity,
homogeneity and translatability properties of multidimensional
conditional $g$-expectations and multidimensional dynamic risk
measures; we prove that a multidimensional dynamic $g$-risk measure
is nonincreasingly convex if and only if the generator $g$ satisfies
a quasi-monotone increasingly convex condition. A general dual
representation is given for the multidimensional dynamic convex
$g$-risk measure in which the penalty term is expressed more
precisely. It is shown that model uncertainty leads to the convexity of risk measures.
As to applications, we show how this
multidimensional approach can be applied to measure the insolvency risk of a firm with interacted
subsidiaries; optimal risk sharing for $\protect\gamma $-tolerant $g$-risk
measures is investigated. Insurance $g$-risk measure and other ways to induce $g$-risk measures are also studied at the end
of the paper.
\end{abstract}


\end{frontmatter}

\section{INTRODUCTION}

Quantifying the risk of the uncertainty in the future value of a portfolio
is one of the key tasks of risk management. This quantification is usually
achieved by modeling the uncertain payoff as a random variable, to which
then a certain functional is applied. Such functionals are usually called
risk measures. Coherent risk measures have been introduced by \textrm{\cite%
{adeh1,adeh2}}, and further developed by \textrm{\cite{d1,d2}}; sublinear
risk measures by \textrm{\cite{f1}}; convex risk measures by \textrm{\cite%
{fs1,fs2}} and \textrm{\cite{fr1}}. In all these papers, static risk
measures are under consideration. However practically we are interested in
monitoring the riskiness of a position $\xi $ at any intermediate time $t$
between the initial date $0$ and the final $T$. The dynamical risk measure
represents the riskiness at time $t$, taking into account all the
information available up to date $t$. A class of dynamic risk measures,
called $g$-risk measure, can be obtained by means of conditional $g$%
-expectations (\textrm{\cite{jl,ro}}). In particular, any dynamic $g$-risk
measure satisfies the \textquotedblleft time-consistency
property\textquotedblright . Note that the notion of time consistency first
appeared in \textrm{\cite{p2}} under the Brownian filtration. See \textrm{%
\cite{ks}, \cite{bn08a, bn09a}} and references therein for more recent
papers dealing with dynamic risk measures on a filtered probability space.
Particularly \textrm{\cite{bn08a, bn09a}} characterized the time consistency
by a \textquotedblleft cocycle condition\textquotedblright\ for the minimal
penalty function of a convex risk measure. And using the theory of BMO%
\footnote{%
The space of functions of bounded mean oscillation.} martingales, she also
provided a new class of time consistent risk measures even with jumps. A
characterization of time consistency for a concave monetary utility function
in terms of its acceptance set was presented in \textrm{\cite{cdk}}. Further
discussions on dynamic risk measures can be found in \textrm{\cite{be}},
\textrm{\cite{fr2}}, \textrm{\cite{jr}}, \textrm{\cite{rse}}, among many
others. The variety and complexity of the many risk factors and their
interaction give rise to the need for a multidimensional approach to risk
measurement. The nature of complex marketable assets has increased the
requirements for necessary analytical methods. These methods must deal with
the multiplicity of risk sources and their correlations. Many economists and
mathematical finance scholars have attempted to achieve a multidimensional
framework for risk analysis (\textrm{\cite{fv,jmt,ku}}). The notion of
multidimensional static risk measure was introduced in \textrm{\cite{jmt}}
and \textrm{\cite{ku}}. The objective of the paper is to deal with
multidimensional dynamic risk measures induced by multidimensional
conditional $g$-expectations. To obtain a preliminary intuition of $g$-risk
measures, we give a brief introduction. Let $\left( Y^{1},\ldots
,Y^{n}\right) $ be the unique solution to multidimensional backward
stochastic differential equation (BSDE for short) \eqref{2.1}. $\left(
Y^{1},\ldots ,Y^{n}\right) $ is also called multidimensional conditional $g$%
-expectation here, denoted by $\left( \mathcal{E}_{g_{1}}^{1}[\cdot |%
\mathcal{F}_{t}],\ldots ,\mathcal{E}_{g_{n}}^{n}[\cdot |\mathcal{F}%
_{t}]\right) $. The $g$-risk measure is then defined as $\left( \rho
_{t}^{g_{1}}[\xi ^{1}],\ldots ,\rho _{t}^{g_{n}}[\xi ^{n}]\right) =\left(
\mathcal{E}_{g_{1}}^{1}[-\xi ^{1}|\mathcal{F}_{t}],\ldots ,\mathcal{E}%
_{g_{n}}^{n}[-\xi ^{n}|\mathcal{F}_{t}]\right) $. If $\xi ^{k},k=1,\ldots n,$
represents the potential loss of the $k$th subsidiary of a firm, then $\rho
_{t}^{g_{k}}[\xi ^{k}]$ calculates its risk at time $t$. Due to the
interaction among subsidiaries, $\rho _{t}^{g_{1}}[\xi ^{1}],\ldots ,\rho
_{t}^{g_{n}}[\xi ^{n}]$ are related to each other through the generator ($%
g_{1},\ldots g_{n}$). See the application section for more intuitive
examples. Note that the $g$-risk measure depends heavily on its generator $g$%
. The coefficient $g$ can be interpreted as infinitesimal risk measure over
a time interval $[t,t+dt]$ as pointed out in \textrm{\cite{be}}. The choice
of $g$ is crucial. Particularly, \textrm{\cite{ce} }showed that the function
$g$ could depend on the preferences of investors. For one dimensional case,
based on results in \textrm{\cite{chmp}}, \textrm{\cite{p2}}, \textrm{\cite%
{dpr}},\ authors there showed that almost any dynamic risk measure under a
Brownian filtration (satisfying certain domination condition) can be
represented by a conditional $g$-expectation and an explicit representation
of the penalty term of general dynamic concave utilities with continuity
from above can be obtained by applying the theory of $g$-expectation. This
is also a motivation for us to investigate multidimensional risk measures
induced by multidimensional $g$-expectations.

$g$-expectations which were introduced in \textrm{\cite{p1} }provide an
excellent explanation for the famous Allais's paradox. $g$-expectation is an
extension of the classical mathematical expectation. It preserves much of
the properties of standard mathematical expectation except the linearity.
The notion of $g$-expectation can be applied to describe nonlinear recursive
utilities, risk measures and the pricing of contingent claims, see \textrm{%
\cite{epq}} and references therein for an overview of applications.

As far as now, few papers studied multidimensional $g$-expectations. One of
the important reasons is that we can not give an explicit necessary and
sufficient condition for the comparison theorem of solutions of
multidimensional backward stochastic differential equations, and the form of
the multidimensional comparison theorem is more complicated than the $1$%
-dimensional one. Starting from the point of backward stochastic viability
property (BSVP) in \textrm{\cite{bqr}}, \textrm{\cite{hp}} gave a necessary
and sufficient condition under which the comparison theorem holds for
multidimensional BSDEs. This paper develops an explicit expression for the
necessary and sufficient condition of the multidimensional comparison
theorem which leads the multidimensional comparison theorem to be more
applicable.

Note that the multidimensional $g$-expectation is not a notion just by
putting together several one dimensional $g$-expectations. It has its own
complications due to the high dimensions. For a one dimensional $g$-risk
measure induced by a $g$-expectation, the monotonicity holds naturally due
to the comparison theorem of BSDEs. But for multidimensional $g$-risk
measures we need the \textquotedblleft quasi-monotonicity
condition\textquotedblright\ of the generator $g$ throughout the paper. For
the multidimensional case any axiom is a strong condition. In fact one
should add axioms to multidimensional $g$-risk measures carefully. We will
see later that if one expect to find examples of $g$-risk measures which
satisfy the axioms of monotonicity and translation invariance, one may
discover that the multidimensional $g$-risk measure is in fact $n$ one
dimensional $g$-risk measures put together. We extend notions of convex and
coherent risk measures in this paper by dropping the axiom of translation
invariance.

We also present some applications of multidimensional $g$-risk measures. As
interpreted in \textrm{\cite{km}} and \textrm{\cite{kow}}, economic agents
do, in fact, communicate with each other and learn from each other. They
also infer information from the actions of others and, most importantly, in
most markets they trade with each other. Within a multi-business firm,
subsidiaries interact with each other. If we understand how their mutual
interactions influence the decisions, we can model their insolvency risk as
a multidimensional BSDE. Recall that \textrm{\cite{Jarrow} }measured the
insolvency risk of a firm by the cost of a put option written on the firm's
net value. Such an option guarantees that the value of the firm's assets
will always be greater than or equal to the value of the firm's liabilities.
So the firm could write a set of related put options for her subsidiaries to
protect them against insolvency. The measure defined by the price of the put
option is called \textquotedblleft Put Premium Risk
Measure\textquotedblright . The acceptance set generated by this put premium
risk measure generates a risk measure satisfying the axiom of translation
invariance but may violating time consistency except that the original put
premium risk measure also satisfies the axiom of translation invariance.
This risk measure characterizes the minimal capital required to protect the
firm's subsidiaries against insolvency.

To deduce properties of multidimensional $g$-risk measures, we have to make
some mathematical preliminaries first. Readers who are interested only in
financial applications can jump to section 4 and 5 with little knowledge
about BSDEs. The paper is organized as follows. Some results on backward
stochastic viability property are recalled in section 2. Section 3 contains
the explicit expressions for the multidimensional comparison theorem and
viability on a rectangle. Some basic properties and Jensen's inequality for
multidimensional $g$-expectations are studied in section 4. Section 5 deals
with multidimensional dynamic risk measures induced by multidimensional
conditional $g$-expectations. The multidimensional $g$-risk measure is shown
to be a multidimensional dynamic time consistent risk measure. We prove that
a multidimensional dynamic $g$-risk measure is nonincreasingly convex if and
only if the generator $g$ satisfies a quasi-monotone increasingly convex
condition. A general dual representation is also given for a
multidimensional dynamic convex $g$-risk measure in which the penalty term
is expressed more precisely. An application of our multidimensional approach
is taken into account to measure the insolvency risk of a multi-business
firm with interacted business units. Optimal risk sharing for $\gamma $%
-tolerant $g$-risk measures is investigated. Insurance $g$-risk measure,
multidimensional $g$-pricing mechanism and stochastic differential utility
are also considered.

\section{BACKWARD STOCHASTIC VIABILITY PROPERTY AND MULTIDIMENSIONAL
COMPARISON}

Let $(B_{t})_{t\in \lbrack 0,T]}$ be a standard $d$-dimensional Brownian
motion on a probability space $(\Omega ,\mathcal{F},P)$ and $(\mathcal{F}%
_{t})_{t\in \lbrack 0,T]}$ be the augmented Brownian filtration generated by
$(B_{t})_{t\in \lbrack 0,T]}$. Let $T<0$ be a fixed time. For any Euclidean
space $H$, we define for $p\geq 1$, $L_{\mathcal{F}}^{p}(0,T;H)$:=\{all $H$%
-valued and $\mathcal{F}_{t}$-adapted stochastic processes such that $E\left[
\int_{0}^{T}|\varphi _{t}|^{p}dt\right] <\infty $\}; $S_{\mathcal{F}%
}^{p}(0,T;H)$:=\{all continuous processes in $L_{\mathcal{F}}^{p}(0,T;H)$
such that $E\left[ \sup_{0\leq t\leq T}|\varphi _{t}|^{p}dt\right] <\infty $%
\}; $L^{p}(\mathcal{F}_{T};H)$:=\{the space of all $H$-valued and $\mathcal{F%
}_{T}$-measurable random variables such that $E\left[ |\xi |^{p}\right]
<\infty $\}. We say $x\geq y$, for $x,y\in R^{\mathit{n}}$, if $x_{i}\geq
y_{i}$, for all $i=1,2,...,n$. Let $e_{i},i=1,...,n$, denote the orthogonal
basis of vectors in $R^{\mathit{n}}$. $\forall a\in R^{\mathit{n}}$, $c\in
R^{n\times \mathit{d}}$, we use norms $|a|=\sqrt{\sum_{i=1}^{n}|a_{i}|^{2}}$%
, $|c|=\sqrt{\sum_{i=1}^{n}\sum_{j=1}^{d}|c_{ij}|^{2}}$.

Consider the following $n$-dimensional BSDE:

\begin{equation}
y_{t}=\xi +\int_{t}^{T}g\left( s,y_{s},z_{s}\right)
ds-\int_{t}^{T}z_{s}dB_{s},\ 0\leq t\leq T,  \label{2.1}
\end{equation}%
where the terminal variable $\xi \in L^{2}(\mathcal{F}_{T};R^{n})$ and the
function $g:\Omega \times \lbrack 0,T]\times R^{n}\times R^{n\times \mathit{d%
}}\longmapsto R^{n}$ is progressively measurable. We call $(\xi ,g,T)$ the
standard parameters of BSDE \eqref{2.1}.

For the function $g$, we assume throughout the paper that

(H1) $P$-$a.s.$, $\forall(y,z)$, $t\rightarrow g(t,y,z)$ is continuous;

(H2) There exists a\ constant $L\geqslant0$ such that $P$-$a.s.$, we have: $%
\forall t$, $\forall(y,y^{\prime})$, $\forall(z,z^{\prime})$,
\begin{equation*}
|\mathit{g}\left( t,y,z\right) -\mathit{g}\left( t,y^{\prime},z^{\prime
}\right) |\leq L\left( |y-y^{\prime}|+|z-z^{\prime}|\right) ;
\end{equation*}

(H3) $\sup_{0\leq t\leq T}|g(t,0,0)|\in L^{2}(\mathcal{F}_{T};R^{n}).$

Let $K$ be a nonempty closed convex set of $R^{n}$. We define $\Pi_{K}(a)$
as the projection of point $a$ onto $K,$ and $d_{K}(\cdot)$ the distance
function of $K$. The following proposition comes from \textrm{\cite{bqr}}.
Throughout this paper, it is understood that an equation or an inequality
holds true always in the sense of $P$-$a.s.$.

\begin{proposition}
Suppose that $g$ satisfies (H1) $\thicksim$(H3). Then the following are
equivalent:

(i) For any $t\in \lbrack 0,T]$, $\forall \xi \in L^{2}(\mathcal{F}_{t};K)$,
the unique solution $\left( Y,Z\right) \in S_{\mathcal{F}}^{2}(0,t;R^{n})%
\times L_{\mathcal{F}}^{2}(0,t;R^{n\times d})$ to the following BSDE over
time interval $[0,t]$:
\begin{equation}
y_{s}=\xi +\int_{s}^{t}g\left( r,y_{r},z_{r}\right)
dr-\int_{s}^{t}z_{r}dB_{r},\ 0\leq s\leq t,  \label{2.2}
\end{equation}%
satisfies $Y_{s}\in K$, $\forall s\in \lbrack 0,t],$ $P$-$a.s.$.

(ii) $\forall (t,z)\in \lbrack 0,T]\times R^{n\times d}$ and $\forall y\in
R^{n}$\ such that $d_{K}^{2}(\cdot )$ is twice differentiable at $y$,
\begin{equation}
4\left\langle y-\Pi _{K}(y),g(t,\Pi _{K}(y),z)\right\rangle \leq \
\left\langle D^{2}d_{K}^{2}(y)z,z\right\rangle +Cd_{K}^{2}(y),
\end{equation}%
where $C>0$ is a constant which does not depend on $(t,y,z)$, $D^{2}f(y)$
denotes the second order derivative of the function f.
\end{proposition}

For the following two BSDEs,%
\begin{equation}
y_{s}^{i}=\xi ^{i}+\int_{s}^{t}g^{i}\left( r,y_{r}^{i},z_{r}^{i}\right)
dr-\int_{s}^{t}z_{r}^{i}dB_{r},\ 0\leq s\leq t,\ i=1,2.  \label{2.4}
\end{equation}

\textrm{\cite{hp}} proved

\begin{proposition}
Suppose that $g^{1}$, $g^{2}$ satisfy (H1) $\thicksim$(H3). Then the
following are equivalent:

(i) \ For any $t\in \lbrack 0,T],\ \forall \xi ^{1},\xi ^{2}\in L^{2}(%
\mathcal{F}_{t};R^{n})$\ such that $\xi ^{1}\geq \xi ^{2},$\ the unique
solutions $\left( Y^{1},Z^{1}\right) $ and $\left( Y^{2},Z^{2}\right) $ in $%
S_{\mathcal{F}}^{2}(0,t;R^{n})\times L_{\mathcal{F}}^{2}(0,t;R^{n\times d})$
to BSDEs \eqref{2.4} over time interval $[0,t]$ satisfy:%
\begin{equation*}
\quad Y_{s}^{1}\geq Y_{s}^{2},\ s\in \lbrack 0,t],
\end{equation*}

(ii)\ $\forall t$, $\forall (y,y^{\prime })$, $\forall (z,z^{\prime })$,%
\begin{equation}
-4\left\langle y^{-},g^{1}(t,y^{+}+y^{\prime },z)-g^{2}(t,y^{\prime
},z^{\prime })\right\rangle \ \leq \ 2\sum_{k=1}^{n}I_{y_{k}<0}\left\vert
z_{k}-z_{k}^{\prime }\right\vert ^{2}+C\left\vert y^{-}\right\vert ^{2},\
\label{2.5}
\end{equation}%
where $C>0$ is a constant.
\end{proposition}

We could not get an obvious intuition of a greater-than or less-than
relation between $g^{1}$and $g^{2}$ through \eqref{2.5}. An explicit
expression of \eqref{2.5} is given in the next section just like the well
known one dimensional comparison theorem, see Theorem 2.2 in \textrm{\cite%
{epq}}.

\section{EXPLICIT EXPRESSION FOR MULTIDIMENSIONAL COMPARISON THEOREM AND
VIABILITY ON A RECTANGLE}

\subsection{Comparison theorem}

This subsection presents a criterion equivalent to condition \eqref{2.5}
given by \textrm{\cite{hp}.} This criterion, of simpler form than Hu and
Peng's condition has been exploited as a sufficient condition to prove the
multidimensional comparison theorem by \textrm{\cite{zhb}} and \textrm{\
\cite{zsw}}. By a short and forward calculus, we show that Hu and Peng's
criterion coincides with condition in \textrm{\cite{zhb}}.

\begin{theorem}
Suppose that $g^{1}$, $g^{2}$ satisfy (H1) $\thicksim$(H3). Then the
following are equivalent:

(i) \ For any $t\in \lbrack 0,T],\ \forall \xi ^{1},\xi ^{2}\in L^{2}(%
\mathcal{F}_{t};R^{n})$,\ such that $\xi ^{1}\geq \xi ^{2},$\ the unique
solutions $\left( Y^{1},Z^{1}\right) $ and $\left( Y^{2},Z^{2}\right) $ in $%
S_{\mathcal{F}}^{2}(0,t;R^{n})\times L_{\mathcal{F}}^{2}(0,t;R^{n\times d})$
to BSDEs \eqref{2.4} over time interval $[0,t]$ satisfy,%
\begin{equation*}
\quad Y_{s}^{1}\geq Y_{s}^{2},\ s\in \lbrack 0,t],
\end{equation*}

(ii)\ For any k=1,2,...,n, $\forall t\in \lbrack 0,T]$, $\forall y^{\prime
}\in R^{n}$,%
\begin{equation}
g_{k}^{1}(t,\delta ^{k}y+y^{\prime },z)\geq g_{k}^{2}(t,y^{\prime
},z^{\prime }).  \label{3.1}
\end{equation}%
for any $\delta ^{k}y\in R^{n}$ such that $\delta ^{k}y\geq 0$, $(\delta
^{k}y)_{k}=0$,\footnote{%
Here we follow notations in \textrm{\cite{hp} }for easy comparison. We
denote by $\delta ^{k}y$ a constant belonging to $R^{n}$ associated with $%
g_{k}$ and $(\delta ^{k}y)_{k}$, the $k$th component of $\delta ^{k}y$.} $%
z,z^{\prime }\in R^{n\times d}$ and $(z)_{k}=(z^{\prime })_{k}$.
\end{theorem}

\textbf{Proof.}$\ $ \ From Proposition 2.2, (i) is equivalent to
\begin{equation}
-4\left\langle y^{-},g^{1}(t,y^{+}+y^{\prime },z)-g^{2}(t,y^{\prime
},z^{\prime })\right\rangle \ \leq \ 2\sum_{k=1}^{n}I_{y_{k}<0}\left\vert
z_{k}-z_{k}^{\prime }\right\vert ^{2}+C\left\vert y^{-}\right\vert ^{2}.
\label{3.2}
\end{equation}%
Putting in \eqref{3.2} $y=\delta ^{k}y+y_{k}\cdot e_{k},$ where $y_{k}\in
R^{-}$, $\delta ^{k}y\in R^{n}$, $\delta ^{k}y\geq 0$, $(\delta ^{k}y)_{k}=0$%
, we get

\begin{equation*}
4y_{k}(g_{k}^{1}(t,\delta ^{k}y+y^{\prime },z)-g_{k}^{2}(t,y^{\prime
},z^{\prime }))\leq \ 2\left\vert z_{k}-z_{k}^{\prime }\right\vert
^{2}+C\left\vert y_{k}\right\vert ^{2}.
\end{equation*}%
Taking $z_{k}=z_{k}^{\prime }$, dividing $4y_{k}$ and letting $%
y_{k}\rightarrow 0^{-}$, we deduce that $g_{k}^{1}(t,\delta ^{k}y+y^{\prime
},z)\geq g_{k}^{2}(t,y^{\prime },z^{\prime })$.

Conversely, From assumption (H2) and condition (ii), we have $\forall t\in
\lbrack 0,T]$, $\forall y^{\prime }\in R^{n}$, $\forall z,z^{\prime }\in
R^{n\times d}$ and $z^{\prime \prime }=(z_{1}^{\prime },z_{2}^{\prime
},...z_{k-1}^{\prime },z_{k},z_{k+1}^{\prime },...,z_{n}^{\prime })$,
\begin{align*}
g_{k}^{1}(t,\delta ^{k}y+y^{\prime },z)-g_{k}^{2}(t,y^{\prime },z^{\prime
})& =g_{k}^{1}(t,\delta ^{k}y+y^{\prime },z)-g_{k}^{2}(t,y^{\prime
},z^{\prime \prime })+g_{k}^{2}(t,y^{\prime },z^{\prime \prime
})-g_{k}^{2}(t,y^{\prime },z^{\prime }) \\
& \geq 0-L\left\vert z_{k}-z_{k}^{\prime }\right\vert .
\end{align*}%
Then we have for $y=y_{k}\cdot e_{k},$ $y_{k}\in R$, $y_{k}<0$,
\begin{align}
4y_{k}(g_{k}^{1}(t,\delta ^{k}y+y^{\prime },z)-g_{k}^{2}(t,y^{\prime
},z^{\prime }))& \leq \ -4Ly_{k}\left\vert z_{k}-z_{k}^{\prime }\right\vert
\leq 2\left\vert z_{k}-z_{k}^{\prime }\right\vert ^{2}+2L^{2}\left\vert
y_{k}\right\vert ^{2}  \notag \\
& =\ 2I_{y_{k}<0}\left\vert z_{k}-z_{k}^{\prime }\right\vert
^{2}+2L^{2}\left\vert y^{-}\right\vert ^{2},  \label{3.3}
\end{align}%
when getting the second inequality, we use $2ab\leq a^{2}+b^{2}$. From %
\eqref{3.3} we deduce easily that
\begin{equation*}
-4\left\langle y^{-},g^{1}(t,y^{+}+y^{\prime },z)-g^{2}(t,y^{\prime
},z^{\prime })\right\rangle \leq \ 2\sum_{k=1}^{n}I_{y_{k}<0}\left\vert
z_{k}-z_{k}^{\prime }\right\vert ^{2}+C\left\vert y^{-}\right\vert ^{2}.
\end{equation*}%
$\square $

By Theorem 3.1 and its proof , we have the following corollary which has
been given in \textrm{\cite{hp}}.

\begin{corollary}
Suppose that $g^{1}=g^{2}=g$ satisfies (H1) $\thicksim$(H3). Then the
following are equivalent:

(i) \ For any $t\in \lbrack 0,T],\ \forall \xi ^{1},\xi ^{2}\in L^{2}(%
\mathcal{F}_{t};R^{n})$,\ such that $\xi ^{1}\geq \xi ^{2},$\ the unique
solutions $\left( Y^{1},Z^{1}\right) $ and $\left( Y^{2},Z^{2}\right) $ in $%
S_{\mathcal{F}}^{2}(0,t;R^{n})\times L_{\mathcal{F}}^{2}(0,t;R^{n\times d})$
to BSDEs \eqref{2.4} over time interval $[0,t]$ satisfy:%
\begin{equation*}
\quad Y_{s}^{1}\geq Y_{s}^{2},\ s\in \lbrack 0,t],
\end{equation*}%
$\quad $(ii)\ For any $k=1,2,...,n$, for any $t$ and $y^{\prime }$, $g_{k}$
does not depend on $\left( z_{j}\right) _{j\neq k}$ and%
\begin{equation}
g_{k}(t,\delta ^{k}y+y^{\prime },z_{k})\geq g_{k}(t,y^{\prime },z_{k}),
\label{3.4}
\end{equation}%
for any $\delta ^{k}y\in R^{n}$ such that $\delta ^{k}y\geq 0$, $(\delta
^{k}y)_{k}=0$.
\end{corollary}

\textbf{Proof.}$\ $ In \eqref{3.1}, let $\delta ^{k}y=0$, $%
z=(z_{1},...,z_{n})$, $z^{\prime }=(0,...,z_{k},...,0)$, and interchange
positions of $z^{\prime }$ and $z$, we can deduce that for any $t$ and $y$, $%
g_{k}$ does not depend on $\left( z_{j}\right) _{j\neq k}$. Then the
equivalency of (i) and (ii) is a consequence of Theorem 3.1. $\square $

\begin{remark}
If condition \eqref{3.4} holds for $g$, we say that $g(t,\cdot ,z)$ is
quasi-monotonously increasing. The \textquotedblleft quasi-monotonously
increasing\textquotedblright\ \ condition was first introduced for
stochastic differential equations by \textrm{\cite{mel} }and later cited by
\textrm{\cite{gm}}, \textrm{\cite{dw}} to prove a comparison theorem for
multidimensional stochastic differential equations.

When condition \eqref{3.4} holds for g, by a direct calculation, the
Lipschitz condition
\begin{equation*}
\left\vert {}g_{k}\left( s,y,z_{k}\right) -{}g_{k}\left( s,y^{\prime
},z_{k}^{\prime }\right) \right\vert \leq \mu \left( \left\vert
y-{}y^{\prime }\right\vert +\left\vert z_{k}-z_{k}^{\prime }\right\vert
\right)
\end{equation*}
is equivalent to
\begin{eqnarray*}
-\mu \left( \sum_{l\neq k}\left( y_{l}-{}y_{l}^{\prime }\right)
^{-}+\left\vert y_{k}-{}y_{k}^{\prime }\right\vert +\left\vert
z_{k}-z_{k}^{\prime }\right\vert \right) &\leq &{}g_{k}\left(
s,y,z_{k}\right) -{}g_{k}\left( s,y^{\prime },z_{k}^{\prime }\right) \\
&\leq &\mu \left( \sum_{l\neq k}\left( y_{l}-{}y_{l}^{\prime }\right)
^{+}+\left\vert y_{k}-{}y_{k}^{\prime }\right\vert +\left\vert
z_{k}-z_{k}^{\prime }\right\vert \right) ,
\end{eqnarray*}%
for each $k=1,\ldots ,n$.
\end{remark}

\subsection{Uniqueness theorem}

For two multidimensional BSDEs, if their generators coincide, of course they
have the same solution; conversely if for any same terminal datum, they
always have the same solution, do their generators coincide? We give an
affirmative answer in Theorem 3.2.

\begin{proposition}
Suppose that $g$ satisfies (H1) $\thicksim$(H3). Then the following are
equivalent:

(i) \ For any $t\in \lbrack 0,T]$,$\ \forall \xi ^{1},\xi ^{2}\in L^{2}(%
\mathcal{F}_{t};R^{n})$\ such that $(\xi ^{1}-\xi ^{2})\in K$,\ the unique
solutions $\left( Y^{1},Z^{1}\right) $ and $\left( Y^{2},Z^{2}\right) $ in $%
S_{\mathcal{F}}^{2}(0,t;R^{n})\times L_{\mathcal{F}}^{2}(0,t;R^{n\times d})$
to BSDEs \eqref{2.4} over time interval $[0,t]$ satisfy:%
\begin{equation*}
\quad (Y_{s}^{1}-Y_{s}^{2})\in K,\ \forall s\in \lbrack 0,t],
\end{equation*}%
$\quad $

(ii)\ $\forall t$, $\forall (y,y^{\prime })$, $\forall (z,z^{\prime })$,%
\begin{equation}
4\left\langle y-\Pi _{K}(y),g^{1}(t,\Pi _{K}(y)+y^{\prime
},z)-g^{2}(t,y^{\prime },z^{\prime })\right\rangle \leq \ \left\langle
D^{2}d_{K}^{2}(y)(z-z^{\prime }),z-z^{\prime }\right\rangle +Cd_{K}^{2}(y),
\label{3.6}
\end{equation}%
where $C>0$ is a constant which does not depend on $(t,y,z)$, $D^{2}f(y)$
denotes the second order derivative of the function f.
\end{proposition}

\textbf{Proof.}$\ \ \ $ Consider the following BSDE over time interval $%
[0,t] $, \
\begin{equation}
\widetilde{Y}_{s}=\widetilde{\xi }+\int_{s}^{t}\widetilde{g}\left( r,%
\widetilde{Y}_{r},\widetilde{Z}_{r}\right) dr-\int_{s}^{t}\widetilde{Z}%
_{r}dB_{r},  \label{3.7}
\end{equation}%
where for $\widetilde{y}=(\widetilde{y}^{1},\widetilde{y}^{2})$, $\widetilde{%
z}=(\widetilde{z}^{1},\widetilde{z}^{2})$,
\begin{equation*}
\widetilde{g}\left( r,\widetilde{y},\widetilde{z}\right) =\left( g^{1}(r,%
\widetilde{y}^{1}+\widetilde{y}^{2},\widetilde{z}^{1}+\widetilde{z}%
^{2})-g^{2}(r,\widetilde{y}^{2},\widetilde{z}^{2}),g^{2}(r,\widetilde{y}^{2},%
\widetilde{z}^{2})\right) .
\end{equation*}

Set $\widetilde{Y}_{s}=(Y_{s}^{1}-Y_{s}^{2},Y_{s}^{2})$, $\widetilde{Z}%
_{s}=(Z_{s}^{1}-Z_{s}^{2},Z_{s}^{2})$. Then (i) is equivalent to the
following:

(iii) For any $t\in \lbrack 0,T]$,$\ \forall \widetilde{\xi }=(\widetilde{%
\xi }^{1},\widetilde{\xi }^{2})$ such that $\widetilde{\xi }^{1}\in L^{2}(%
\mathcal{F}_{t};K)$,\ the unique solutions $\left( \widetilde{Y}_{s},%
\widetilde{Z}_{s}\right) $ in $S_{\mathcal{F}}^{2}(0,t;R^{n})\times L_{%
\mathcal{F}}^{2}(0,t;R^{n\times d})$ to BSDE \eqref{3.7} over time interval $%
[0,t]$ satisfy:%
\begin{equation*}
\quad \widetilde{Y}_{s}^{1}\in K,\ \forall s\in \lbrack 0,t].
\end{equation*}

Applying Proposition 2.1 to BSDE \eqref{3.7} and the convex closed set $%
K\times R^{n}$, obviously (iii) is equivalent to (ii). $\square $

As a consequence of Proposition 3.1, we have the following uniqueness result
for generators of multidimensional BSDEs \eqref{2.4}.

\begin{theorem}
For any $t\in \lbrack 0,T],\ \forall \xi ^{1}=\xi ^{2}\in L^{2}(\mathcal{F}%
_{t};R^{n})$,%
\begin{equation*}
\quad Y_{s}^{1}=Y_{s}^{2},\ \forall s\in \lbrack 0,t],
\end{equation*}%
\ if and only if \ for any $k=1,2,...,n$, $\forall (t,y,z)\in \lbrack
0,T]\times R^{n}\times R^{n\times d}$,%
\begin{equation*}
g_{k}^{1}(t,y,z)=g_{k}^{2}(t,y,z).
\end{equation*}
\end{theorem}

\textbf{Proof.}$\ \ \ $ Take $K=\{0\}$, by Proposition 3.1, It is seen that
\textquotedblleft$Y^{1}=Y^{2}$\textquotedblright\ is equivalent to

\begin{equation}
4\left\langle y,g^{1}(t,y^{\prime },z)-g^{2}(t,y^{\prime },z^{\prime
})\right\rangle \leq \ 2\sum_{k=1}^{n}\left\vert z_{k}-z_{k}^{\prime
}\right\vert ^{2}+C\left\vert y\right\vert ^{2}.  \label{3.8}
\end{equation}%
Putting in \eqref{3.8} $y=y_{k}\cdot e_{k}$, where $y_{k}\in R$, $y_{k}>0$, $%
z=z^{\prime }$, we deduce that for any $k=1,2,...,n$, $\forall (t,y,z)\in
\lbrack 0,T]\times R^{n}\times R^{n\times d}$, $g_{k}^{1}(t,y,z)\leq
g_{k}^{2}(t,y,z)$. Similarly the case \textquotedblleft $y_{k}<0$%
\textquotedblright\ leads to that \textquotedblleft $g^{1}\geq g^{2}$%
\textquotedblright . Thus we obtain the desired result. $\square $

\begin{remark}
Comparison Theorem 3.1 does not imply the above uniqueness result. If one of
$Y^{i}$ is increasing in $\xi ^{i}$, then it just implies that\
\textquotedblleft $\xi ^{1}=\xi ^{2},Y^{1}=Y^{2}$\ $\Leftrightarrow $ $%
g_{k}^{1}(t,y,z_{k})=g_{k}^{2}(t,y,z_{k})$\textquotedblright , in which for
any $(t,y)$, $g_{k}$ does not depend on $\left( z_{j}\right) _{j\neq k}$,
due to the \textquotedblleft quasi-monotonicity condition\textquotedblright\ %
\eqref{3.4}. Note that in Theorem 3.2, g could depend on $\left(
z_{1},\ldots ,z_{n}\right) $.
\end{remark}

\subsection{Viability on a rectangle}

Rectangles are special closed convex subset of $R^{n}$. We first give
explicit expressions for necessary and sufficient conditions for
non-negative and non-positive solutions of BSDEs.

\begin{theorem}
Suppose that $g$ satisfies (H1) $\thicksim$(H3). Then the following are
equivalent:

(i) \ For any $t\in \lbrack 0,T],\ \forall \xi \in L^{2}(\mathcal{F}%
_{t};R^{n})$,\ such that $\xi \geq 0$, ($\xi \leq 0$),\ the unique solution $%
\left( Y,Z\right) $ over time interval $[0,t]$ satisfies,%
\begin{equation*}
Y_{s}\geq 0,\ (Y_{s}\leq 0),s\in \lbrack 0,t],
\end{equation*}

(ii)\ For any $k=1,2,...,n$, $\forall t\in \lbrack 0,T]$,%
\begin{equation}
g_{k}(t,\delta ^{k}y,\delta ^{k}z)\geq 0,\ (g_{k}(t,-\delta ^{k}y,\delta
^{k}z)\leq 0)  \label{3.9}
\end{equation}%
for any $\delta ^{k}y\in R^{n}$ such that $\delta ^{k}y\geq 0$, $(\delta
^{k}y)_{k}=0$, and $\delta ^{k}z\in R^{n\times d}$, $(\delta ^{k}z)_{k}=0$.
\end{theorem}

\textbf{Proof.}$\ $ \ If (i) holds, then from Proposition 2.1, we have%
\begin{equation}
-4\left\langle y^{-},g(t,y^{+},z)\right\rangle \ \leq \
2\sum_{k=1}^{n}I_{y_{k}<0}\left\vert z_{k}\right\vert ^{2}+C\left\vert
y^{-}\right\vert ^{2}.  \label{3.10}
\end{equation}%
Take $y=\delta ^{k}y+y_{k}\cdot e_{k}$ in \eqref{3.10}$,$ $y_{k}\in R^{-}$, $%
\delta ^{k}y\in R^{n}$, $\delta ^{k}y\geq 0$, $(\delta ^{k}y)_{k}=0$, we get%
\begin{equation*}
4y_{k}g_{k}(t,\delta ^{k}y,z)\leq \ 2\left\vert z_{k}\right\vert
^{2}+C\left\vert y_{k}\right\vert ^{2},
\end{equation*}%
dividing $4y_{k}$ and letting $\ z_{k}=0$, $y_{k}\rightarrow 0^{-}$, we
deduce that $g_{k}(t,\delta ^{k}y,\delta ^{k}z)\geq 0$.

Conversely, if \eqref{3.9} holds, by assumption (H2), we have $\forall t\in
\lbrack 0,T]$,%
\begin{equation*}
g_{k}(t,\delta ^{k}y,z)-g_{k}(t,\delta ^{k}y,\delta ^{k}z)\geq -L\left\vert
z_{k}\right\vert ,
\end{equation*}%
thus $g_{k}(t,\delta ^{k}y,z)\geq g_{k}(t,\delta ^{k}y,\delta
^{k}z)-L\left\vert z_{k}\right\vert \geq -L\left\vert z_{k}\right\vert $.
Then for $y_{k}\in R$, $y_{k}<0$,%
\begin{equation*}
4y_{k}(g_{k}(t,\delta ^{k}y,z)\leq \ -4Ly_{k}\left\vert z_{k}\right\vert
\leq 2\left\vert z_{k}\right\vert ^{2}+2L^{2}\left\vert y_{k}\right\vert
^{2}=\ 2I_{y_{k}<0}\left\vert z_{k}\right\vert ^{2}+2L^{2}\left\vert
y^{-}\right\vert ^{2},
\end{equation*}%
Thus we can get easily that$-4\left\langle y^{-},g(t,y^{+},z)\right\rangle
\leq \ 2\sum_{k=1}^{n}I_{y_{k}<0}\left\vert z_{k}\right\vert
^{2}+C\left\vert y^{-}\right\vert ^{2}$.

The result in the case \textquotedblleft $\xi \leq 0$\textquotedblright\ can
be deduced from the preceding case \textquotedblleft $\xi \geq 0$%
\textquotedblright\ applied to the well chosen function $\overline{g}%
(t,y,z):=-g(t,-y,-z)$, $\forall (t,y,z)\in \lbrack 0,T]\times R^{n}\times
R^{n\times d}$. \medskip$\square $

Minus $C\in R^{n}$ on both sides of BSDE \eqref{2.2}, we have
\begin{equation}
y_{s}-C=\xi -C+\int_{s}^{t}g\left( r,y_{r}-C+C,z_{r}\right)
dr-\int_{s}^{t}z_{r}dB_{r},\ 0\leq s\leq t,
\end{equation}%
let $\overline{Y}_{s}=Y_{s}-C,\overline{\xi }=\xi -C$, $\overline{g}%
(r,y,z)=g(r,y+C,z)$, then apply Theorem 3.3 to the following

\begin{equation}
\overline{y}_{s}=\overline{\xi }+\int_{s}^{t}\overline{g}\left( r,\overline{y%
}_{r},\overline{z}_{r}\right) dr-\int_{s}^{t}\overline{z}_{r}dB_{r},\ 0\leq
s\leq t,
\end{equation}%
we have

\begin{corollary}
\ Suppose that $g$ satisfies (H1) $\thicksim $(H3).\ For any $t\in \lbrack
0,T],\ $\

(i) \ Assume $C\in R^{n}$. $\forall \xi \in L^{2}(\mathcal{F}_{t};R^{n})$,\
such that $\xi \geq C$,\ ($\xi \leq C$),\ then
\begin{equation*}
\quad Y_{s}\geq C,\ (Y_{s}\leq C),\forall s\in \lbrack 0,t],
\end{equation*}%
if and only if for any $k=1,2,...,n$, $\forall t\in \lbrack 0,T]$,
\begin{equation}
g_{k}(t,\delta ^{k}y+C,\delta ^{k}z)\geq 0,\ (g_{k}(t,-\delta ^{k}y+C,\delta
^{k}z)\leq 0)
\end{equation}%
for any $\delta ^{k}y\in R^{n}$ such that $\delta ^{k}y\geq 0$, $(\delta
^{k}y)_{k}=0$, and $\delta ^{k}z\in R^{n\times d}$, $(\delta ^{k}z)_{k}=0$.

(ii)\ Assume condition \eqref{3.4} and $C_{1},C_{2}\in R^{n}$. $\forall \xi
\in \lbrack C_{1},C_{2}]$,%
\begin{equation*}
\quad Y_{s}\in \lbrack C_{1},C_{2}],\forall s\in \lbrack 0,t],
\end{equation*}%
if and only if for any $k=1,2,...,n$, $\forall t\in \lbrack 0,T]$,
\begin{equation}
g_{k}(t,\delta ^{k}y+C_{1},z_{k}=0)\geq 0\geq g_{k}(t,-\delta ^{k}y^{\prime
}+C_{2},z_{k}^{\prime }=0),  \label{3.14}
\end{equation}%
for any $\delta ^{k}y,\delta ^{k}y^{\prime }\in R^{n}$ such that $\delta
^{k}y,\delta ^{k}y^{\prime }\in \lbrack 0,C_{2}-C_{1}]$, $(\delta
^{k}y)_{k}=(\delta ^{k}y^{\prime })_{k}=0$. $\square $
\end{corollary}

\begin{remark}
Assuming condition\ \eqref{3.4}, let $\delta ^{k}y=\delta ^{k}y^{\prime }=0$%
, $C_{1}=C_{2}$ in \eqref{3.14}, we can deduce that $\forall \xi =C$, $C$ is
a constant in $R^{n}$,
\begin{equation*}
\quad Y_{s}=C,s\in \lbrack 0,t],
\end{equation*}%
if and only if for any $k=1,2,...,n$, $\forall t\in \lbrack 0,T]$, for any $%
t $ and $C$,
\begin{equation}
g_{k}(t,C,z_{k}=0)=0.
\end{equation}%
In fact the condition \eqref{3.4} is not necessary for the constancy (see
Theorem 4.1(i) in Section 4).
\end{remark}

\section{MULTIDIMENSIONAL $g$-EXPECTATIONS}

\subsection{Basic properties}

A BSDE is in fact a dynamical mechanism of nonlinear expectation. It is
natural to define multidimensional $g$-expectation by multidimensional BSDEs.

\begin{definition}
For each $\xi \in L^{2}(\mathcal{F}_{t};R^{n})$, let (Y, Z) be the solution
of BSDE \eqref{2.2}, we define multidimensional g-expectation by $\mathcal{E}%
_{g}^{t}[\xi ]=Y_{0}$, multidimensional conditional $g$-expectation by $%
\mathcal{E}_{g}^{t}[\xi |\mathcal{F}_{s}]=Y_{s}$, $s\in \lbrack 0,t].$
\end{definition}

The following properties hold for the multidimensional conditional $g$%
-expectation.

\begin{theorem}
Suppose that $g$ satisfies (H1) $\thicksim $(H3).\ For all $t\in \lbrack
0,T] $,$\ $\

(i)\ Constancy. $\forall C\in R^{n}$,%
\begin{equation*}
\mathcal{E}_{g}^{t}[C|\mathcal{F}_{s}]=C,\forall s\in \lbrack 0,t],
\end{equation*}%
if and only if for any $k=1,2,...,n$, $\forall t\in \lbrack 0,T]$,%
\begin{equation}
g_{k}(t,C,z=0)=0,  \label{4.1}
\end{equation}

(ii) Monotonicity.\ For all $\xi ^{1},\xi ^{2}\in L^{2}(\mathcal{F}%
_{t};R^{n})$\ such that $\xi ^{1}\geq \xi ^{2},$\
\begin{equation*}
\quad \mathcal{E}_{g}^{t}[\xi ^{1}|\mathcal{F}_{s}]\geq \mathcal{E}%
_{g}^{t}[\xi ^{2}|\mathcal{F}_{s}],\ \forall s\in \lbrack 0,t],
\end{equation*}%
if and only if for any $k=1,2,...,n$, $\forall t\in \lbrack 0,T]$, $\forall
y^{\prime }\in R^{n}$, $g_{k}$ does not depend on $\left( z_{j}\right)
_{j\neq k}$ and
\begin{equation*}
g_{k}(t,\delta ^{k}y+y^{\prime },z_{k})\geq g_{k}(t,y^{\prime },z_{k}).
\end{equation*}%
for any $\delta ^{k}y\in R^{n}$ such that $\delta ^{k}y\geq 0$, $(\delta
^{k}y)_{k}=0$.

(iii) Positivity.\ $\forall \xi \in L^{2}(\mathcal{F}_{t};R^{n})$\ such that
$\xi \geq 0$,\ then
\begin{equation*}
\quad \mathcal{E}_{g}^{t}[\xi |\mathcal{F}_{s}]\geq 0,\forall s\in \lbrack
0,t],
\end{equation*}%
if and only if for any $k=1,2,...,n$, $\forall t\in \lbrack 0,T]$,
\begin{equation*}
g_{k}(t,\delta ^{k}y,\delta ^{k}z)\geq 0,
\end{equation*}%
for any $\delta ^{k}y\in R^{n}$ such that $\delta ^{k}y\geq 0$, $(\delta
^{k}y)_{k}=0$, and $\delta ^{k}z\in R^{n\times d}$, $(\delta ^{k}z)_{k}=0$.

(iv) Time-consistency. For any $0\leq s\leq t\leq T$,
\begin{equation*}
\quad \mathcal{E}_{g}^{T}[\xi |\mathcal{F}_{s}]=\mathcal{E}_{g}^{t}[\mathcal{%
E}_{g}^{T}[\xi |\mathcal{F}_{t}]|\mathcal{F}_{s}],\xi \in L^{2}(\mathcal{F}%
_{T};R^{n}).
\end{equation*}

(v)\ Homogeneity. $\forall \xi \in L^{2}(\mathcal{F}_{t};R^{n})$, $\forall
a\in R$,
\begin{equation*}
\mathcal{E}_{g}^{t}[a\xi |\mathcal{F}_{s}]=a\mathcal{E}_{g}^{t}[\xi |%
\mathcal{F}_{s}],\forall s\in \lbrack 0,t],
\end{equation*}%
if and only if for all $k=1,2,...,n$, $\forall t\in \lbrack 0,T]$,
\begin{equation*}
g_{k}(t,ay,az)=ag_{k}(t,y,z),
\end{equation*}%
for all $(y,z)\in R^{n}\times R^{n\times d}$.
\end{theorem}

\textbf{Proof.}$\ $ \ (i) The sufficiency is obvious. We now prove the
necessity. Let $(Y_{s})$ be the first part of solution $(Y,Z)$ of
\begin{equation}
\qquad Y_{s}=C+\int_{s}^{t}g\left( r,Y_{r},Z_{r}\right)
dr-\int_{s}^{t}Z_{r}dB_{r},\ 0\leq s\leq t,  \label{4.2}
\end{equation}%
Minus $C\in R^{n}$ on both sides of BSDE \eqref{4.2} and let $\overline{Y}%
_{s}=Y_{s}-C$, then $(\overline{Y}_{s},\overline{Z}_{s})$ satisfies the
following
\begin{equation*}
\overline{Y}_{s}=0+\int_{s}^{t}g\left( r,\overline{Y}_{r}+C,\overline{Z}%
_{r}\right) dr-\int_{s}^{t}\overline{Z}_{r}dB_{r},\ 0\leq s\leq t.
\end{equation*}%
Take $K=\{0\}$, then by Proposition 2.1, $\forall \xi =0$, $\overline{Y}%
_{s}=0$, $\forall s\in \lbrack 0,t]$ if and only if $\forall (t,y,z)$,
\begin{equation}
4\left\langle y,g(t,C,z)\right\rangle \leq \ 2\sum_{k=1}^{n}\left\vert
z_{k}\right\vert ^{2}+C\left\vert y\right\vert ^{2}.  \label{4.3}
\end{equation}%
Let $z=0$, (coordinate-wise), $y=y_{k}\cdot e_{k}$ in \eqref{4.3}$,$ $%
y_{k}\in R^{+}$, and $y_{k}\rightarrow 0^{+}$, we get that for any $%
k=1,2,...,n$, $\forall t\in \lbrack 0,T]$, $g_{k}(t,C,0)\leq 0$; if $y_{k}<0$%
, and $y_{k}\rightarrow 0^{-}$, we have $g_{k}(t,C,0)\geq 0$. Thus for any $%
k=1,2,...,n$, $\forall t\in \lbrack 0,T]$, $g_{k}(t,C,0)=0$.

(ii) and (iii) are consequences of Corollary 3.1 and Theorem 3.3
respectively. The time consistency (iv) is due to the uniqueness of solution
of BSDE \eqref{2.2}.

To prove (v), applying Theorem 3.2 to the following two BSDEs:

\begin{equation*}
\overline{Y}_{s}=\xi +\int_{s}^{t}\frac{1}{a}g\left( r,a\overline{Y}_{r},a%
\overline{Z}_{r}\right) dr-\int_{s}^{t}\overline{Z}_{r}dB_{r},\ a\neq 0,s\in
\lbrack 0,t],
\end{equation*}

\begin{equation*}
Y_{s}=\xi +\int_{s}^{t}g\left( r,Y_{r},Z_{r}\right)
dr-\int_{s}^{t}Z_{r}dB_{r},\ s\in \lbrack 0,t],
\end{equation*}%
we obtain that for any $\xi \in L^{2}(\mathcal{F}_{t};R^{n})$,\ $\overline{Y}%
_{s}=\frac{1}{a}\mathcal{E}_{g}^{t}[a\xi |\mathcal{F}_{s}]=\mathcal{E}%
_{g}^{t}[\xi |\mathcal{F}_{s}]=Y_{s}$, $\forall s\in \lbrack 0,t]$ if and
only if $g(t,ay,az)=ag(t,y,z)$. When $a=0$, it is just the constancy. $%
\square $

\begin{remark}
The constancy in Theorem 4.1 is for all constant in $R^{n}$. In fact for any
fixed $C\in R^{n}$, we still have $\mathcal{E}_{g}^{t}[C|\mathcal{F}%
_{s}]=C,\forall s\in \lbrack 0,t]$ if and only if for any $k=1,2,...,n$, $%
\forall s\in \lbrack 0,T]$, $g_{k}(t,C,0)=0$. The proof is analogous. For
instance, given $C\in R^{n}$, a g-expectation with generator $%
g:=r(y-C)+\theta z$ has the property of constancy for only $C$.
\end{remark}

\begin{remark}
We do not impose in advance special conditions on multidimensional
g-expectations in order to explore the general relations between
g-expectations and their generating mechanism g, which is good for showing
the essence of properties for multidimensional g-expectations. As we see
above, multidimensional g-expectations are more complicated than the
1-dimensional one because they are related to different dimensions.
\end{remark}

\subsection{Jensen's inequality}

Let $\varphi =(\varphi _{1},\ldots ,\varphi _{n})$ be a multiple convex
function, i.e., $\varphi _{k}:R\rightarrow R$, $k=1,2,...,n$, is convex: $%
\forall x,y\in R$, $\forall \lambda \in \left[ 0,1\right] $, $\varphi
_{k}(\lambda x+\left( 1+\lambda y\right) )\leq \lambda \varphi
_{k}(x)+\left( 1+\lambda \right) \varphi _{k}(y)$. This section shows that
if Jensen's inequality holds for a multidimensional conditional $g$%
-expectation, then the $g$-expectation consists in fact of $n$ one
dimensional $g$-expectations with generator $g$ satisfying positively
homogeneous and negatively super-homogeneous conditions w.r.t. $z$.

\begin{definition}
Jensen's inequality for multidimensional conditional g-expectation holds, if
for any $t\in \lbrack 0,T]$, for all $\xi \in L^{2}(\mathcal{F}_{t};R^{n})$,
convex function $\varphi :R^{n}\rightarrow R^{n}$ such that $\varphi (\xi
)\in L^{2}(\mathcal{F}_{t};R^{n})$, and for any $k=1,2,...,n$, a.s., $%
\forall s\in \lbrack 0,t]$,%
\begin{equation}
\mathcal{E}_{g_{k}}^{t}[\varphi _{k}(\xi _{k})|\mathcal{F}_{s}]\geq \varphi
_{k}(\mathcal{E}_{g_{k}}^{t}[\xi _{k}|\mathcal{F}_{s}]).  \label{4.4}
\end{equation}
\end{definition}

Let $\xi _{k}$ be the payoff of a derivative at time $T$. Let $\mathcal{E}%
_{g_{k}}^{T}[\xi _{k}]$ be its price at time 0. The above Jensen's
inequality means that a convex transformation of the payoff $\xi _{k}$ at
time $T$ yields a higher price than the convex transformation of the initial
price. So traders in financial markets should take care of this difference.
For instance, consider a European call option with payoff $\max \left[
\left( S_{T}-L\right) ,0\right] $, where $S_{T}$ is the price of the
underlying stock at time $T$, $L$ is the strike price. If Jensen's
inequality holds, then $\mathcal{E}_{g}[\max \left[ \left( S_{T}-L\right) ,0%
\right] ]\geq \max [\mathcal{E}_{g}\left[ S_{T}-L\right] ,0]$, which means
that the price of the call option is greater than the nonnegative value
obtained by a direct calculation of $\left( S_{T}-L\right) $.

Jensen's inequality\ is in fact a comparison for solutions of two BSDEs. Let
$\varphi _{k}(x)=x+a,x\in R,a\in R^{+}$. From the Jensen's inequality we get
that $\mathcal{E}_{g_{k}}^{t}[\xi _{k}+a|\mathcal{F}_{s}]\geq \mathcal{E}%
_{g_{k}}^{t}[\xi _{k}|\mathcal{F}_{s}]+a\geq \mathcal{E}_{g_{k}}^{t}[\xi
_{k}|\mathcal{F}_{s}]$, $\forall k=1,2,...,n$, a.s., $\forall s\in \lbrack
0,t]$. So it is natural to require the \textquotedblleft quasi-monotonicity
condition\textquotedblright\ \eqref{3.4} for the generator $g$.

\begin{remark}
Jensen's inequality can be used to evaluate the nonlinear expectation
operator $\mathcal{E}_{g}^{t}$. A familiar example is $\mathcal{E}%
_{g_{k}}^{t}[\left\vert \xi _{k}\right\vert |\mathcal{F}_{s}]\geq \left\vert
\mathcal{E}_{g_{k}}^{t}[\xi _{k}|\mathcal{F}_{s}]\right\vert $, for any $%
k=1,2,...,n$, a.s., $\forall s\in \lbrack 0,T]$. For a nonlinear expectation
$\mathcal{E}_{g}^{t}$, we have to impose additional conditions on $g$ to
obtain Jensen's inequality.
\end{remark}

\begin{theorem}
Suppose that $g$ satisfies (H1) $\thicksim $(H3) and condition \eqref{3.4}.
Jensen's inequality for a multidimensional conditional g-expectation holds
if and only if

(i)\ for any $k=1,2,...,n$, $g_{k}$ does not depend on y and $\forall
(t,z_{k})\in\lbrack0,T]\times R^{d}$,

$\ \ g_{k}(t,\lambda z_{k})=\lambda g_{k}(t,z_{k})$, $\forall\lambda\geq0$; $%
g_{k}(t,\lambda z_{k})\geq\lambda g_{k}(t,z_{k})$, $\forall\lambda<0$.
\end{theorem}

\textbf{Proof.}$\ $ \ If condition (i) holds, the multidimensional $g$%
-expectation is in fact $n$ one dimensional $g$-expectation. Then by
analogous proof of Jensen's inequality for one dimensional $g$-expectation
(see \textrm{\cite{jc}}), we can obtain the multidimensional Jensen's
inequality.

We wish to emphasize some interesting aspects above the converse. Assume
that Jensen's inequality for a multidimensional conditional $g$-expectation
holds. For any $y\in R^{n}$, take $\varphi _{y}(x)=x+y$, $x\in R^{n}$. From
the Jensen's inequality and the \textquotedblleft
monotonicity\textquotedblright\ of $\mathcal{E}_{g}^{t}[\cdot \,|\mathcal{F}%
_{s}]$, we have $\forall s\in \lbrack 0,t]$, for any $\delta \xi \in L^{2}(%
\mathcal{F}_{t};R_{+}^{n})$,
\begin{equation*}
\mathcal{E}_{g}^{t}[\xi +\delta \xi +y|\mathcal{F}_{s}]\geq \mathcal{E}%
_{g}^{t}[\xi |\mathcal{F}_{s}]+y.
\end{equation*}%
Then respectively $\overline{Y}_{s}:=\mathcal{E}_{g}^{t}[\xi +\delta \xi +y|%
\mathcal{F}_{s}]-y$ and $Y_{s}:=\mathcal{E}_{g}^{t}[\xi |\mathcal{F}_{s}]$
are solutions of
\begin{equation}
\overline{Y}_{s}=\xi +\delta \xi +\int_{s}^{t}g\left( r,\overline{Y}_{r}+y,%
\overline{Z}_{r}\right) dr-\int_{s}^{t}\overline{Z}_{r}dB_{r},\ 0\leq s\leq
r.  \label{4.5}
\end{equation}%
\begin{equation}
Y_{s}=\xi +\int_{s}^{t}g\left( r,Y_{r},Z_{r}\right)
dr-\int_{s}^{t}Z_{r}dB_{r},\ 0\leq s\leq t,  \label{4.6}
\end{equation}%
For BSDE \eqref{4.5} and BSDE \eqref{4.6}, by Comparison Theorem 3.1, we
have for any $k=1,2,...,n$, $\forall t\in \lbrack 0,T]$, for any $\overline{y%
},y\in R^{n}$, $z_{k}\in R^{d}$, $\delta ^{k}y\in R^{n}$ and $\delta
^{k}y\geq 0$, $(\delta ^{k}y)_{k}=0$,
\begin{equation}
g_{k}\left( t,\overline{y}+y+\delta ^{k}y,z_{k}\right) \geq g_{k}\left( t,%
\overline{y},z_{k}\right) .  \label{4.7}
\end{equation}%
Similarly taking $\varphi _{-y}(x)=x-y$, then we can also get that
\begin{equation}
g_{k}\left( t,\overline{y}-y+\delta ^{k}y,z_{k}\right) \geq g_{k}\left( t,%
\overline{y},z_{k}\right) .  \label{4.8}
\end{equation}%
From \eqref{4.7} and \eqref{4.8}, it is easy to deduce that $g\left(
t,y,z_{k}\right) =g\left( t,0,z_{k}\right) $. Therefore $g$ does not depend
on $y$.

Take $\varphi _{\lambda }(x)=\lambda x$, $\lambda \in R\backslash \{0\},x\in
R^{n}$. From Jensen's inequality and the monotonicity of $\mathcal{E}%
_{g}^{t}[\cdot \,|\mathcal{F}_{s}]$, we have $\forall s\in \lbrack 0,t],$ $%
\forall \lambda \in R\backslash \{0\}$, for any $\delta \xi \in L^{2}(\Omega
,\mathcal{F}_{t},P;R_{+}^{n})$,
\begin{equation*}
\mathcal{E}_{g}^{t}[\lambda \xi +\delta \xi |\mathcal{F}_{s}]\geq \lambda
\mathcal{E}_{g}^{t}[\xi |\mathcal{F}_{s}].
\end{equation*}%
Then $\overline{Y}_{s}:=\mathcal{E}_{g}^{t}[\lambda \xi +\delta \xi |%
\mathcal{F}_{s}]$ and $Y_{s}:=\lambda \mathcal{E}_{g}^{t}[\xi |\mathcal{F}%
_{s}]$ are solutions of
\begin{equation}
\overline{Y}_{s}=\lambda \xi +\delta \xi +\int_{s}^{t}g\left( r,\overline{Z}%
_{r}\right) dr-\int_{s}^{t}\overline{Z}_{r}dB_{r},\ 0\leq s\leq t.
\label{4.9}
\end{equation}%
\begin{equation}
Y_{s}=\lambda \xi +\int_{s}^{t}\lambda g\left( r,\frac{1}{\lambda }%
Z_{r}\right) dr-\int_{s}^{t}Z_{r}dB_{r},\ 0\leq s\leq r,  \label{4.10}
\end{equation}%
For BSDE \eqref{4.9} and BSDE \eqref{4.10}, by Comparison Theorem 3.1, we
have for any $k=1,2,...,n$, $\forall t\in \lbrack 0,T]$, $\forall \lambda
\in R\backslash \{0\}$,
\begin{equation}
g_{k}(t,\lambda z_{k})\geq \lambda g_{k}(t,z_{k}).  \label{4.11}
\end{equation}%
When $\lambda \in (0,1]$, we have%
\begin{equation}
\frac{1}{\lambda }g_{k}(t,z_{k})\geq g_{k}(t,\frac{1}{\lambda }z_{k}).
\end{equation}%
Thus $\forall \lambda \in \lbrack 1,\infty )$,
\begin{equation}
\lambda g_{k}(t,z_{k})\geq g_{k}(t,\lambda z_{k}).  \label{4.13}
\end{equation}%
Then combining \eqref{4.13} with \eqref{4.11}, we get that $\forall \lambda
\in \lbrack 1,\infty )$, $g_{k}(t,\lambda z_{k})=\lambda g_{k}(t,z_{k})$.
When $\lambda \in (0,1)$, the positive homogeneity holds too.

Let $z_{k}=0$ in \eqref{4.11}, we get $g_{k}(t,0)\geq \lambda g_{k}(t,0)$, $%
\forall \lambda \in R\backslash \{0\}$. Therefore for any $k=1,2,...,n$, $%
\forall t\in \lbrack 0,T]$,%
\begin{equation}
g_{k}(t,0)\geq -g_{k}(t,0),  \label{4.14}
\end{equation}%
and%
\begin{equation}
g_{k}(t,0)\geq 2g_{k}(t,0).  \label{4.15}
\end{equation}%
\eqref{4.14} and \eqref{4.15} lead to that for any $k=1,2,...,n$, $\forall
t\in \lbrack 0,T]$, $g_{k}(t,0)=0$. $\square $

\begin{remark}
The above theorem shows that if Jensen's inequality holds for a
multidimensional g-risk measure (up to a minus sign, nonlinear expectation),
then this risk measure is not recursive (g does not depend on y) in the
generating mechanism\ g and different dimensions are not interacted with
each other. The risk could be measured dimension by dimension respectively.
\end{remark}

We refer \textrm{\cite{jp}} for Jensen's inequality of $g$-convex function
under one dimensional $g$-expectation. $g$-convexity is a new notion of
convex functions. They study this problem from a different point of view:
for a given generator $g$, to characterize the function $\varphi $ (not
necessarily convex) such that the Jensen's inequality holds. For Jensen's
inequality \eqref{4.4} and Jensen's inequality in \textrm{\cite{jc}}, we aim
to characterize the generator $g$ such that the Jensen's inequality holds
for all convex functions. Note also that a convex function is not
necessarily a $g$-convex function.

\section{MULTIDIMENSIONAL DYNAMIC RISK MEASURES}

Since Artzner et al.'s seminal work on coherent risk measures, see \textrm{%
\cite{adeh1, adeh2}}, many authors are interested in the problem of how to
measure the risk of a financial position as mentioned in the introduction.
However they discover that the classical risk measure -VaR, could not
satisfy some natural time-consistency. The notion of time-consistency was
first proposed by \textrm{\cite{p2}} for nonlinear expectations (up to a
sign, nonlinear risk measures) in the framework of Brownian filtration.
Recently a lot of papers concern time-consistent dynamic risk measures, see
\textrm{\cite{bn08a,bn09a}, \cite{cdk}, \cite{ks}} for characterizations of
time consistency on a general filtered probability space; see \textrm{\cite%
{p2}, \cite{ro}, \cite{jl}} for dynamic $g$-risk measures introduced by
BSDEs; see also \textrm{\cite{be}, \cite{jr} }and \textrm{\cite{rse}} for
time consistent risk measures. This section defines a multidimensional risk
measure by multidimensional $g$-expectation and shows its properties and
applications.

Let $(\Omega ,\left( \mathcal{F}_{t}\right) _{t\in \lbrack 0,T]},P)$ be a
filtered probability space. Let $\mathcal{G}_{T}$ be the space of all
financial positions with maturity $T$, i.e., the set of $R^{n}$-valued $%
\mathcal{F}_{T}$-measurable functions defined on $\Omega $. An element of $%
\mathcal{G}_{T}$ may be the net worth at maturity $T$ of a financial
contract, or the accounting value of a firm's equity, or the surplus of an
insurance company.

\begin{definition}
A multidimensional static risk measure is a functional $\rho :\mathcal{G}_{T}%
\mathcal{\mapsto }R^{n}$ such that (i) $\rho \lbrack \xi ^{1}]\leq \rho
\lbrack \xi ^{2}]$, if $\forall \xi ^{1}\geq \xi ^{2}\in \mathcal{G}_{T}$;
(ii)\ $\rho \lbrack \mathbf{0}]=\mathbf{0}$. The quantity $\rho \lbrack \xi
] $ represents the riskiness of financial position $\xi \in \mathcal{G}_{T}$%
. $\xi $ is acceptable if $\rho \lbrack \xi ]\leq 0$, and unacceptable
otherwise.

We call $\rho _{s,t}:\mathcal{G}_{t}\mathcal{\mapsto }L^{2}(\mathcal{F}%
_{s};R^{n})$, $0\leq s\leq t\leq T$, a multidimensional dynamic risk measure
if

({\normalsize A1}) Monotonicity. $\forall s\in \lbrack 0,t]$, $\rho
_{s,t}[\xi ^{1}]\leq \rho _{s,t}[\xi ^{2}]$, if $\forall \xi ^{1}\geq \xi
^{2}\in \mathcal{G}_{t}$;

({\normalsize A2}) $\rho _{t,t}[\xi ]=-\xi $, $\forall \xi \in \mathcal{G}%
_{t}$;

({\normalsize A3}) Time-consistency. $\mathcal{\rho }_{s,T}[\xi ]=\mathcal{%
\rho }_{s,t}[-\mathcal{\rho }_{t,T}[\xi ]]$, $\forall \xi \in \mathcal{G}%
_{T} $, $\forall 0\leq s\leq t\leq T$;

({\normalsize A4}) Regularity. $\mathbf{1}_{A}\mathcal{\rho }_{s,t}[\xi
\mathbf{1}_{A}]=\mathbf{1}_{A}\mathcal{\rho }_{s,t}[\xi ]$, $\forall \xi \in
\mathcal{G}_{t}$, $\forall A\in \mathcal{F}_{s}$;

({\normalsize A5}) Normalization. $\rho _{s,t}[\mathbf{0}]=\mathbf{0}$.
\end{definition}

Condition (A5) means that \textquotedblleft zero position, zero
risk\textquotedblright . \textrm{\cite{ro} }interpreted the time consistency
as: computing the static risk measure $\mathcal{\rho }[\xi ]$ directly is
equivalent to calculating $\mathcal{\rho }[\xi ]$ in two steps, i.e.
evaluating first the riskiness $\mathcal{\rho }_{t,T}[\xi ]$ of $\xi $ at an
intermediate date $t$ and then quantifying at time $0$ the risk of $-%
\mathcal{\rho }_{t,T}[\xi ]$ through $\mathcal{\rho }$. The financial
meaning of monotonicity is clear: The downside risk of a position is reduced
if the payoff profile is increased. In a sense, the additional information
available to the agent has to be fully used when assessing the riskiness of
a payoff $\xi $, which means in particular that if we know that certain
event $A\in \mathcal{F}_{s}$ is prevailing, then the riskiness of $\xi $
should depend on only what is really possible to happen, i.e. on the
restriction of $\xi $ to $A$. This is characterized by ({\normalsize A4}).
The property ({\normalsize A4}) is equivalent to: $\mathcal{\rho }_{s,t}[\xi
^{1}\mathbf{1}_{A}+\xi ^{2}\mathbf{1}_{A^{c}}]=\mathcal{\rho }_{s,t}[\xi
^{1}]\mathbf{1}_{A}+\mathcal{\rho }_{s,t}[\xi ^{2}]\mathbf{1}_{A^{c}}$, $%
\forall A\in \mathcal{F}_{s}$. Also, ({\normalsize A4}) plus ({\normalsize A5%
}) implies $\mathcal{\rho }_{s,t}[\xi \mathbf{1}_{A}]=\mathbf{1}_{A}\mathcal{%
\rho }_{s,t}[\xi ]$, $\forall \xi \in \mathcal{G}_{t}$, $\forall A\in
\mathcal{F}_{s}$. We refer to \textrm{\cite{ds}} and \textrm{\cite{p2}} for
the proofs.

\subsection{Multidimensional $g$-risk measure}

We call the generator $g$ risk mechanism. In the sequence we consider $%
\mathcal{G}_{t}\mathcal{=}L^{2}(\mathcal{F}_{t};R^{n})$.

\begin{definition}
Suppose that the risk mechanism $g$ satisfies (H1) $\thicksim $(H3).\ For
any $t\in \lbrack 0,T]$,$\ $the risk position $\xi \in L^{2}(\mathcal{F}%
_{t};R^{n})$,\ we define $\rho _{0,t}^{g}[\xi ]=\mathcal{E}_{g}^{t}[-\xi ]$,
$\rho _{s,t}^{g}[\xi ]=\mathcal{E}_{g}^{t}[-\xi |\mathcal{F}_{s}]$, $s\in
\lbrack 0,t]$, as the multidimensional static $g$-risk measure and
multidimensional dynamic $g$-risk measure respectively. Without ambiguity,
we denote $\rho ^{g}=\rho _{0,T}^{g}$, $\rho _{t}^{g}=\rho _{t,T}^{g}$.
\end{definition}

The financial meaning of $\rho _{t}^{g}[\xi ]$ is: Let $\xi $ be the
negative part of an uncertain income at time $T$, then $-\xi $ represents
the potential loss at time $T$. Mathematically speaking, losses are viewed
as negative gains. The $g$-expectation is in fact a pricing operator (%
\textrm{\cite{p3})}. Therefore $\mathcal{E}_{g}^{T}[-\xi |\mathcal{F}_{t}]$
is the discount value at time $t$ of potential loss $-\xi $. It is the risk
for an investor. The $k$th component $\rho _{t}^{g_{k}}[\xi ^{k}]$ of $\rho
_{t}^{g}[\xi ]=\left( \rho _{t}^{g_{1}}[\xi ^{1}],\ldots ,\rho
_{t}^{g_{n}}[\xi ^{n}]\right) $ represents the risk of the $k$th subsidiary
of a firm. $\rho _{t}^{g_{1}}[\xi ^{1}],\ldots ,\rho _{t}^{g_{n}}[\xi ^{n}]$
are related to each other through the generator ($g_{1},\ldots g_{n}$). The
link between measures of risk and BSDEs is particularly interesting because
it enhances interpretation and tractability of risk measures. \textrm{\cite%
{be}} point out that the coefficient $g$ of BSDEs can be interpreted as
infinitesimal risk measure over a time interval $[t,t+dt]$ as $E[dY_{t}|%
\mathcal{F}_{t}]=-g(t,Y_{t},Z_{t})dt$ where $Z_{t}$ is the local volatility
of the conditional risk measure, $V(dY_{t}|\mathcal{F}_{t})=|Z_{t}|^{2}dt$.
Choosing carefully the coefficient $g$ enables to generate dynamic $g$-risk
measures that are locally compatible with the views and practice of
different agents in the market.

For multidimensional $g$-risk measures, additional condition has to be
imposed to insure the monotonicity. Even for the convex and coherent risk
measures in the next section, the quasi-monotonicity condition \eqref{3.4}
is indispensable. It is easy to prove

\begin{theorem}
Suppose that the risk mechanism $g$ satisfies (H1) $\thicksim $(H3) and %
\eqref{3.4} and $g(\cdot ,\mathbf{0},\mathbf{0})=\mathbf{0}$.\ Then $\rho
^{g}[\cdot ]$ and $\rho _{s.t}^{g}[\cdot ]$ are multidimensional static risk
measure and multidimensional dynamic risk measure respectively.
\end{theorem}

\textbf{Proof.}$\ $ \ Theorem 4.1 yields that $\rho ^{g}[\xi ]$ is a
multidimensional static risk measure and $\rho _{s,t}^{g}[\xi ]$ satisfies
(A1) $\thicksim $(A3) and (A5). We now prove (A4). Let $(Y_{s},Z_{s})$, $(%
\overline{Y}_{s},\overline{Z}_{s})$ be solutions of the following BSDEs%
\begin{equation}
Y_{s}=-\xi +\int_{s}^{t}g\left( r,Y_{r},Z_{r}\right)
dr-\int_{s}^{t}Z_{r}dB_{r},\ 0\leq s\leq t,
\end{equation}%
\begin{equation}
\overline{Y}_{s}=-\xi 1_{A}+\int_{s}^{t}g\left( r,\overline{Y}_{r},\overline{%
Z}_{r}\right) dt-\int_{s}^{t}\overline{Z}_{r}dB_{r},\ 0\leq s\leq t.
\end{equation}%
Multiply $1_{A}$ on both sides of the above two BSDEs, since $\forall \left(
y,z\right) \in R^{n}\times R^{n\times d}$,
\begin{equation}
1_{A}g(r,y,z)=g(r,y1_{A},z1_{A}),\forall A\in \mathcal{F}_{s},
\end{equation}%
we get%
\begin{equation}
Y_{s}1_{A}=-\xi 1_{A}+\int_{s}^{t}g\left( r,Y_{r}1_{A},Z_{r}1_{A}\right)
dr-\int_{s}^{t}Z_{r}1_{A}dB_{r},\ 0\leq s\leq t,
\end{equation}%
\begin{equation}
\overline{Y}_{s}1_{A}=-\xi 1_{A}+\int_{s}^{t}g\left( r,\overline{Y}_{r}1_{A},%
\overline{Z}_{r}1_{A}\right) dt-\int_{s}^{t}\overline{Z}_{r}1_{A}dB_{r},\
0\leq s\leq t.
\end{equation}%
Comparing the above two BSDEs, by the uniqueness of solution of BSDE ($-\xi
1_{A},g,t$), we obtain $Y_{s}1_{A}=\overline{Y}_{s}1_{A}$, thus $1_{A}%
\mathcal{\rho }_{s,t}[\xi \mathbf{1}_{A}]=\mathbf{1}_{A}\mathcal{\rho }%
_{s,t}[\xi ]$, $\forall A\in \mathcal{F}_{s}$. The proof is complete. $%
\square $

\subsection{Multidimensional convex and coherent risk measures}

\begin{definition}
A multidimensional dynamic $g$-risk measure $\rho _{s,t}^{g}[\cdot ]$ is
convex if for any $t\in \lbrack 0,T]$, and for all risk positions $\xi
^{1},\xi ^{2}\in L^{2}(\mathcal{F}_{t};R^{n})$, $\forall \lambda \in \lbrack
0,1]$, $\forall s\in \left[ 0,t\right] $, $\rho _{s,t}^{g}[\lambda \xi
^{1}+(1-\lambda )\xi ^{2}]\leq \lambda \rho _{s,t}^{g}[\xi ^{1}]+(1-\lambda
)\rho _{s,t}^{g}[\xi ^{2}].$
\end{definition}

Consider two strategies: one leads to the potential loss $\xi ^{1}$, while
another leads to the potential loss $\xi ^{2}$ . If one diversifies,
spending only the fraction $\lambda $ of the resources on the first strategy
and using the remaining part for the second alternative, the potential loss
is $\lambda \xi ^{1}+(1-\lambda )\xi ^{2}$. Thus, the axiom of convexity
gives a precise meaning to the idea that diversification should not increase
the risk. Monotonicity is a natural condition for one dimensional $g$-risk
measures. Here we still need it for multidimensional $g$-risk measures. For
a convex $g$-risk measure, we have

\begin{theorem}
Suppose that the risk mechanism $g$ satisfies (H1) $\thicksim $(H3).\ For
any $t\in \lbrack 0,T]$, for all risk positions $\xi ^{1},\xi ^{2}\in L^{2}(%
\mathcal{F}_{t};R^{n})$,\ the multidimensional dynamic $g$-risk measure $%
\rho _{s,t}^{g}[\cdot ]$ is nonincreasing and convex if and only if $\forall
t\in \lbrack 0,T]$, $\forall (y^{i},z^{i})\in R^{n}\times R^{n\times d}$, $%
i=1,2,$\ for all $k=1,2,...,n$, $g_{k}$ does not depend on $\left(
z_{j}\right) _{j\neq k}$ and $\forall \lambda \in \lbrack 0,1]$,
\begin{equation}
g_{k}(t,\lambda y^{1}+(1-\lambda )y^{2}-\delta ^{k}y,\ z_{k})\leq \lambda
g_{k}(t,y^{1},z_{k}^{1})+(1-\lambda )g_{k}(t,y^{2},z_{k}^{2}),  \label{5.6}
\end{equation}%
for any $\delta ^{k}y\in R^{n}$ such that $\delta ^{k}y\geq 0$, $(\delta
^{k}y)_{k}=0$ and $z_{k}=\lambda z_{k}^{1}+(1-\lambda )z_{k}^{2}$.
\end{theorem}

\textbf{Proof.}$\ $ Let $(Y_{s}^{1},Z_{s}^{1})$, $(Y_{s}^{2},Z_{s}^{2})$, $(%
\overline{Y}_{s},\overline{Z}_{s})$ be solutions of the following BSDEs%
\begin{equation*}
Y_{s}^{1}=-\xi ^{1}+\int_{s}^{t}g\left( r,Y_{r}^{1},Z_{r}^{1}\right)
dr-\int_{s}^{t}Z_{r}^{1}dB_{r},\ 0\leq s\leq t,
\end{equation*}

\begin{equation*}
Y_{s}^{2}=-\xi ^{2}+\int_{s}^{t}g\left( r,Y_{r}^{2},Z_{r}^{2}\right)
dr-\int_{s}^{t}Z_{r}^{2}dB_{r},\ 0\leq s\leq t,
\end{equation*}%
\begin{equation*}
\overline{Y}_{s}=-\lambda \xi ^{1}-(1-\lambda )\xi ^{2}+\int_{s}^{t}g\left(
r,\overline{Y}_{r},\overline{Z}_{r}\right) dr-\int_{s}^{t}\overline{Z}%
_{r}dB_{r},\ 0\leq s\leq t.
\end{equation*}%
respectively. Set $\widetilde{Y}_{s}=(\lambda Y_{s}^{1}+(1-\lambda
)Y_{s}^{2}-\overline{Y}_{s},\ Y_{s}^{1},\ Y_{s}^{2})$, $\widetilde{Z}%
_{s}=(\lambda Z_{s}^{1}+(1-\lambda )Z_{s}^{2}-\overline{Z}_{s},\ Z_{s}^{1},\
Z_{s}^{2})$. For $y=(y^{1},y^{2},y^{3})\in (R^{n})^{3}$, $%
z=(z^{1},z^{2},z^{3})\in (R^{n\times d})^{3}$, $\lambda \in \lbrack 0,1]$,
we define
\begin{align*}
\widetilde{g}(t,y,z)& =(\lambda g(t,y^{2},z^{2})+(1-\lambda
)g(t,y^{3},z^{3})-g(t,\lambda y^{2}+(1-\lambda )y^{3}-y^{1},\lambda
z^{2}+(1-\lambda )z^{3}-z^{1}), \\
& \ g(t,y^{2},z^{2}),\ g(t,y^{3},z^{3})),
\end{align*}%
Then the convexity and monotonicity of $\rho _{s,t}^{g}[\cdot ]$ is
equivalent to the following

(i) \ For any $t\in \lbrack 0,T]$,$\ \forall \widetilde{\xi }=(\widetilde{%
\xi }^{1},\widetilde{\xi }^{2},\widetilde{\xi }^{3})$ such that $\widetilde{%
\xi }^{1}\geq 0$,\ the unique solutions $\left( \widetilde{Y}_{s},\widetilde{%
Z}_{s}\right) $ in $S_{\mathcal{F}}^{2}(0,t;(R^{n})^{3})\times L_{\mathcal{F}%
}^{2}(0,t;(R^{n\times d})^{3})$ to the following BSDE over time interval $%
[0,t]$:%
\begin{equation}
\widetilde{Y}_{s}=\widetilde{\xi }+\int_{s}^{t}\widetilde{g}\left( r,%
\widetilde{Y}_{r},\widetilde{Z}_{r}\right) du-\int_{s}^{t}\widetilde{Z}%
_{r}dB_{r},\ 0\leq s\leq t.
\end{equation}%
satisfies $\widetilde{Y}\in R_{+}^{n}\times R^{n}\times R^{n}$.

Applying Proposition 2.1 to the above BSDE and the set $R_{+}^{n}\times
R^{n}\times R^{n}$, it is easy to see that (i) is equivalent to

(ii)\ $\forall t$, $\forall(y,y^{\prime})$, $\forall(z,z^{\prime})$,
\begin{align}
& -\left\langle 4y^{-},\lambda
g(t,y^{1},z^{1})+(1-\lambda)g(t,y^{2},z^{2})-g(t,\lambda
y^{1}+(1-\lambda)y^{2}-y^{+},z)\right\rangle  \notag \\
& \leq\ 2\sum_{k=1}^{n}I_{y_{k}<0}\left\vert \lambda z_{k}^{1}+(1-\lambda
)z_{k}^{2}-z_{k}\right\vert ^{2}+C\left\vert y^{-}\right\vert ^{2},
\end{align}
where $C>0$ is a constant which does not depend on $(t,y,z)$.

Similarly to Theorem 3.1, (ii) is equivalent to

(iii) $\forall t\in\lbrack0,T]$, for any $k=1,2,...,n$, $\forall\lambda
\in\lbrack0,1]$,
\begin{equation*}
g_{k}(t,\lambda y^{1}+(1-\lambda)y^{2}-\delta^{k}y,\ z)\leq\lambda
g_{k}(t,y^{1},z^{1})+(1-\lambda)g_{k}(t,y^{2},z^{2}),
\end{equation*}
for any $\delta^{k}y\in R^{n}$ such that $\delta^{k}y\geq0$, $(\delta
^{k}y)_{k}=0$ and $z,z^{1},z^{2}\in R^{n\times d}$ , $z_{k}=\lambda
z_{k}^{1}+(1-\lambda)z_{k}^{2}$.

Let $\lambda =1$, $\delta ^{k}y=0$. we obtain that $g_{k}(t,y^{1},\ z)\leq
g_{k}(t,y^{1},z^{1})$, for any $z,z^{1}\in R^{n\times d}$ such that $%
z_{k}=z_{k}^{1}$. Then it is easy to deduce that $g_{k}$ does not depend on $%
\left( z_{j}\right) _{j\neq k}$. Thus \eqref{5.6} holds true. $\square $

\begin{definition}
A multidimensional dynamic $g$-risk measure $\rho _{s,t}^{g}[\cdot ]$ is
coherent if it satisfies

(i) Positive homogeneity. $\forall \xi \in L^{2}(\mathcal{F}_{t};R^{n})$, $%
\forall a\in R^{+}$, $\rho _{s,t}^{g}[a\xi ]=a\rho _{s,t}^{g}[\xi ],\forall
s\in \lbrack 0,t]$;

(ii) Subadditivity. For all risk positions $\xi ^{1},\xi ^{2}\in L^{2}(%
\mathcal{F}_{t};R^{n})$, $\forall s\in \left[ 0,t\right] $, $\rho
_{s,t}^{g}[\xi ^{1}+\xi ^{2}]\leq \rho _{s,t}^{g}[\xi ^{1}]+\rho
_{s,t}^{g}[\xi ^{2}].$
\end{definition}

For a multidimensional coherent $g$-risk measure, we have

\begin{corollary}
Suppose that the risk mechanism $g$ satisfies (H1) $\thicksim $(H3).\ The
multidimensional dynamic risk measure $\rho _{s,t}^{g}[\cdot ]$ is
nonincreasing and coherent if and only if for all $k=1,2,...,n$, $\forall
t\in \lbrack 0,T]$, $\forall y\in R^{n}$, $g_{k}$ does not depend on $\left(
z_{j}\right) _{j\neq k}$, $\forall a\in R^{+}$,
\begin{equation}
g_{k}(t,ay,az_{k})=ag_{k}(t,y,z_{k}),
\end{equation}%
and $\forall (y^{i},z_{k}^{i})\in R^{n}\times R^{d}$, $i=1,2$,
\begin{equation}
g_{k}(t,y^{1}+y^{2}-\delta ^{k}y,\ z_{k}^{1}+z_{k}^{2})\leq
g_{k}(t,y^{1},z_{k}^{1})+g_{k}(t,y^{2},z_{k}^{2}),
\end{equation}%
for any $\delta ^{k}y\in R^{n}$ such that $\delta ^{k}y\geq 0$, $(\delta
^{k}y)_{k}=0$.
\end{corollary}

\textbf{Proof.}$\ $ Note that a coherent risk measure is equivalent to a
risk measure which is convex and positively homogeneous and for a positively
homogeneous function, it is subadditive if and only if it is convex. Thus
this corollary holds. $\square$

There are examples of dynamic risk measure on filtered probability spaces
which are more general than those generated by a Brownian motion. See
\textrm{\cite{bn08a,bn09a} and \cite{ks}.} Even in the case of Brownian
filtration, $g$-risk measures generated by Lipschitz drivers can not cover
the whole. See the well known entropic risk measure introduced by \textrm{%
\cite{fs2}} and \textrm{\cite{be}} which is characterized by a quadratic
BSDE. There are also coherent risk measure satisfying the axioms of positive
homogeneity and subadditivity but can not be expressed by a $g$-expectation.
\textrm{\cite{p4} }introduced a kind of sublinear expectation, called $G$%
-expectation (capital $G$), via a class of mutually singular probability
measures and found some applications on coherent risk measure. See \textrm{%
\cite{bk}}, \textrm{\cite{ns} }for further study on\textrm{\ }risk measuring
under model uncertainty.

\subsection{Dual representation for multidimensional dynamic convex $g$-risk
measure}

In this section $x\in R^{\mathit{n}}$ means a row vector if without
specification. Let $^{\ast }$ denote the transposition of a vector or a
matrix. For a matrix $A_{m\times m}$, we define $\exp \{{\normalsize A}%
\}=\sum_{n=0}^{\infty }\frac{1}{n!}{\normalsize A}^{n}$. Let $\rho
_{t}^{g}[\cdot ]$ be a convex $g$-risk measure defined in Definition 5.3.
Let $g$ be the generator such that for any $k=1,2,...,n$, $g_{k}$ does not
depend on $\left( z_{j}\right) _{j\neq k}$, quasi-increasingly convex in $%
(y,z_{k})$, i.e. satisfying \eqref{5.6} and $(Y,Z)$ the solution of BSDE
associated to the pair $(\xi ,g)$. We consider the Fenchel-Legendre
transform of $g_{k}$, $k=1,2,...,n$,

\begin{equation*}
G_{k}(t,b,c)=\underset{(y,z)\in R^{n}\times R^{d}}{\sup }[\left\langle
y,b\right\rangle +\left\langle z_{k},c\right\rangle -g_{k}(t,y,z_{k})],
\end{equation*}%
Since $g$ is convex and Lipschitz continuous, we have the duality relation%
\begin{equation}
g_{k}(t,y,z_{k})=\underset{(b,c)\in D_{G}^{(\omega ,t)}}{\sup }[\left\langle
y,b\right\rangle +\left\langle z_{k},c\right\rangle -G_{k}(t,b,c)],
\label{5.11}
\end{equation}%
where $D_{G}:=\left\{ (\omega ,t,b,c)\in \Omega \times \lbrack 0,T]\times
R^{n}\times R^{d}|G(\omega ,t,b,c)<+\infty \right\} $ is the effective
domain of $G$. The $(\omega ,t)$-section of $D_{G}$ is denoted by $%
D_{G}^{(\omega ,t)}$. We denote by $\mathcal{A}$ the set of bounded
progressively measurable processes$(\beta ,\gamma )$ with $(\beta
^{k},\gamma ^{k}),k=1,2,...,n$ valued in $R^{n}\times R^{d}$ such that $%
E\int_{0}^{T}\left\vert G_{k}(t,\beta _{t},\gamma _{t})\right\vert
^{2}dt<\infty $. The boundedness condition on $\mathcal{A}$ means that for
any $(\beta ,\gamma )\in \mathcal{A}$, there exists a constant (dependent of
$(\beta ,\gamma )$) such that $|\beta _{t}^{k}|+|\gamma _{t}^{k}|\leq C$, $t$%
-$a.e,P$-$a.s.$, $k=1,2,...,n$. Let us consider the family of linear
generators

\begin{equation*}
g_{k}^{\beta ,\gamma }(t,y,z_{k})=\left\langle y,\beta _{t}^{k}\right\rangle
+\left\langle z_{k},\gamma _{t}^{k}\right\rangle -G_{k}(t,\beta
_{t}^{k},\gamma _{t}^{k}),\ (\beta ,\gamma )\in \mathcal{A}.
\end{equation*}%
Given $(\beta ,\gamma )\in \mathcal{A}$, we denote by $(Y^{\beta ,\gamma
},Z^{\beta ,\gamma })$ the solution to the following linear BSDE,
\begin{equation}
Y_{t}=-\xi +\int_{t}^{T}\left( Y_{s}\beta _{s}+Z_{s}\gamma _{s}-G(t,\beta
_{s},\gamma _{s})\right) ds-\int_{t}^{T}Z_{s}dB_{s},\ 0\leq t\leq T,
\label{5.12}
\end{equation}%
where we denote $Z=\left(
\begin{array}{c}
Z_{1} \\
\vdots \\
Z_{n}%
\end{array}%
\right) ^{\ast }$; $\beta =\left(
\begin{array}{ccc}
\beta ^{11} & \cdots & \beta ^{n1} \\
\vdots & \cdots & \vdots \\
\beta ^{1n} & \cdots & \beta ^{nn}%
\end{array}%
\right) $; $\gamma ^{i}=\left(
\begin{array}{c}
\gamma ^{i1} \\
\vdots \\
\gamma ^{id}%
\end{array}%
\right) $, $\gamma =\left(
\begin{array}{ccc}
\gamma ^{1} & 0 & 0 \\
0 & \ddots & 0 \\
0 & 0 & \gamma ^{n}%
\end{array}%
\right) $; $Z\gamma =\left(
\begin{array}{cccc}
\left\langle Z_{1},\gamma ^{1}\right\rangle & 0 & \cdots & 0 \\
0 & \left\langle Z_{2},\gamma ^{2}\right\rangle & 0 & \cdots \\
&  & \cdots &  \\
0 & \cdots & 0 & \left\langle Z_{n},\gamma ^{n}\right\rangle%
\end{array}%
\right) $; $\gamma dB=\left(
\begin{array}{cccc}
\left\langle \gamma ^{1},dB\right\rangle & 0 & \cdots & 0 \\
0 & \left\langle \gamma ^{2},dB\right\rangle & 0 & \cdots \\
&  & \cdots &  \\
0 & \cdots & 0 & \left\langle \gamma ^{n},dB\right\rangle%
\end{array}%
\right) $.

\begin{theorem}
Suppose that the risk mechanism $g$ satisfies (H1) $\thicksim $(H3) and
condition \eqref{3.4}.\ The multidimensional dynamic convex risk measure $%
\rho _{t}^{g}[\cdot ]$ over $[0,T]$ has the following representation:

For a risk position $\xi \in L^{2}(\mathcal{F}_{T};R^{n})$,%
\begin{equation}
\rho _{t}^{g}[\xi ]=\mathrm{ess}\underset{(\beta ,\gamma )\in \mathcal{A}}{%
\sup }\left\{ E^{Q^{\gamma }}[-\xi \exp \{\int_{t}^{T}\beta _{u}du\}|%
\mathcal{F}_{t}]-\alpha _{t,T}^{\beta }(Q^{\gamma })\right\} ,
\end{equation}%
where $Q^{\gamma }$ is the probability measure with density process $%
dL_{t}=L_{t}\gamma _{t}dB_{t},L_{0}=\mathbf{1}$ and $\alpha _{t,T}^{\beta
}(Q^{\gamma })=E^{Q^{\gamma }}\left[ \int_{t}^{T}G(s,\beta _{s},\gamma
_{s})\exp \{\int_{t}^{s}\beta _{u}du\}ds|\mathcal{F}_{t}\right] $ is called
the penalty term.
\end{theorem}

\textbf{Proof.}$\ $ Observe from relation \eqref{5.11} and the
\textquotedblleft quasi-monotonicity condition\textquotedblright\ \eqref{3.4}
that for all $(\beta ,\gamma )\in \mathcal{A}$, for any $k=1,2,...,n$, $%
\forall t\in \lbrack 0,T]$, $\forall (y,z_{k})\in R^{n}\times R^{d}$,
\begin{equation}
g_{k}(t,\delta ^{k}y+y,z_{k})\geq g_{k}^{\beta ,\gamma }(t,y,z_{k}).
\end{equation}%
for any $\delta ^{k}y\in R^{n}$ such that $\delta ^{k}y\geq 0$, $(\delta
^{k}y)_{k}=0$. Then by Comparison Theorem 3.1, $Y_{t}\geq Y_{t}^{\beta
,\gamma }$, $\forall t\in \lbrack 0,T]$. Furthermore, since $G$ is convex
with a linear growth condition on its effective domain, for each $(t,\omega
,y,z)$, the supremum in the relation \eqref{5.11} is attained at $(%
\widetilde{b}(t,y,z),\widetilde{c}(t,y,z))$ belonging to the subdifferential
of $-g$, and so $(|\widetilde{b}|,\left\vert \widetilde{c}\right\vert )$ is
bounded by the Lipschitz constant of $g$. By a measurable selection theorem
(see e.g. Appendix in Chapter III of \textrm{\cite{dm}}), since $Y,Z$ are
progressively measurable, we may find a pair of bounded progressively
measurable processes $(\widetilde{\beta },\widetilde{\gamma })$ such that
for any $k=1,2,...,n$, $\forall t\in \lbrack 0,T]$, $\forall (y,z_{k})\in
R^{n}\times R^{d}$,%
\begin{equation}
g_{k}(t,y,z_{k})=g_{k}^{\widetilde{\beta },\widetilde{\gamma }%
}(t,y,z_{k})=\left\langle y,\widetilde{\beta }_{t}^{k}\right\rangle
+\left\langle z_{k},\widetilde{\beta }_{t}^{k}\right\rangle -G_{k}(t,%
\widetilde{\beta }_{t}^{k},\widetilde{\gamma }_{t}^{k}).
\end{equation}%
Then by the Uniqueness Theorem 3.2, these two solutions coincide, and so%
\begin{equation*}
\underset{(\beta ,\gamma )\in \mathcal{A}}{\sup }Y_{t}^{\beta ,\gamma }\geq
Y_{t}^{\widetilde{\beta },\widetilde{\gamma }}=Y_{t}\geq \underset{(\beta
,\gamma )\in \mathcal{A}}{\sup }Y_{t}^{\beta ,\gamma },\forall t\in \lbrack
0,T].
\end{equation*}%
Then by solving the linear BSDE \eqref{5.12}, we end the proof. $\square $

The elements of $\mathcal{A}$ can be interpreted as possible probabilistic
models, which are taken more or less seriously according to the size of the
penalty $\alpha ^{\beta }(Q^{\gamma })$. Thus, the value $\rho _{t}^{g}[\xi
] $ is computed as the worst-case expectation taken over all models $%
Q^{\gamma }$ and penalized by $\alpha ^{\beta }(Q^{\gamma })$. Note that
each $Q^{\gamma }$ is equivalent to the probability measure $P$ with
Radon-Nykodim derivative $\frac{dQ^{\gamma }}{dP}|\mathcal{F}_{t}=\exp \{-%
\frac{1}{2}\int_{0}^{t}\gamma _{u}\gamma _{u}^{\ast }du+\int_{0}^{t}\gamma
_{u}dB_{u}\}$. See \textrm{\cite{dpr}} for the representation of the penalty
term of one dimensional dynamic concave utilities by the theory of one
dimensional $g$-expectation. The representation of coherent $g$-risk measure
is a particular case of the above representation since it corresponds to the
conjugacy function $G=0$ or $+\infty $.

\begin{corollary}
Suppose that the risk mechanism $g$ satisfies (H1) $\thicksim $(H3) and
condition \eqref{3.4}.\ For the multidimensional dynamic coherent risk
measure $\rho _{t}^{g}[\cdot ]$ over $[0,T]$, for a risk position $\xi \in
L^{2}(\Omega ,\mathcal{F}_{T},P;R^{n})$, the following representation holds:%
\begin{equation}
\rho _{t}^{g}[\cdot ]=\mathrm{ess}\underset{(\beta ,\gamma )\in \mathcal{A}}{%
\sup }E^{Q^{\gamma }}\left[ -\xi \exp \{\int_{t}^{T}\beta _{u}du\}|\mathcal{F%
}_{t}\right] .
\end{equation}%
where $Q^{\gamma }$ is the probability measure with density process $%
dL_{t}=L_{t}\gamma _{t}dB_{t},L_{0}=\mathbf{1}$.
\end{corollary}

\subsection{Cash additive risk measures}

\begin{definition}
A multidimensional dynamic $g$-risk measure $\rho _{s,t}^{g}[\cdot ]$ is
called cash additive risk measure if it satisfies the translation
invariance:
\begin{equation*}
\mathcal{\rho }_{s,t}^{g}[\xi +C]=\mathcal{\rho }_{s,t}^{g}[\xi ]-C,\forall
t\in \lbrack 0,t]\text{, }\forall C\in R\mathbf{,}\forall \xi \in L^{2}(%
\mathcal{F}_{t};R^{n}).
\end{equation*}
\end{definition}

\begin{theorem}
Suppose that the risk mechanism $g$ satisfies (H1) $\thicksim $(H3).\ The
multidimensional dynamic $g$-risk measure $\rho _{t}^{g}[\cdot ]$ is cash
additive if and only if for all $k=1,2,...,n$, $\forall t\in \lbrack 0,T]$, $%
\forall z\in R^{n\times d}$, $g$ does not depend on $y$.
\end{theorem}

\textbf{Proof}.$\ $ Set $\overline{Y}_{t}:=\rho _{s,t}^{g}[\xi +C]+C$, $%
Y_{s}:=\rho _{s,t}^{g}[\xi ]$. Then $\overline{Y}_{s}$ and $Y_{s}$ are the
solutions of
\begin{equation}
\overline{Y}_{s}=-\xi +\int_{s}^{t}g\left( r,\overline{Y}_{r}-C,\overline{Z}%
_{r}\right) dr-\int_{s}^{t}\overline{Z}_{r}dB_{r},\ 0\leq s\leq t.
\label{5.15}
\end{equation}%
\begin{equation}
Y_{s}=-\xi +\int_{s}^{t}g\left( r,Y_{r},Z_{r}\right)
dr-\int_{s}^{t}Z_{r}dB_{r},\ 0\leq s\leq t,  \label{5.16}
\end{equation}

For BSDE \eqref{5.15} and BSDE \eqref{5.16}, by the Uniqueness Theorem 3.2,
we obtain the necessary and sufficient condition: for any $k=1,2,...,n$, $%
\forall t\in \lbrack 0,T]$, $\forall (y,z)\in R^{n}\times R^{n\times d}$,
\begin{equation}
g(t,y-C,z)=g(t,y,z).
\end{equation}%
Due to the arbitrariness of $(y,z,C)$, we deduce that $g$ does not depend on
$y$. $\square $

The above cash additivity also holds conditionally: $\forall \xi \in L^{2}(%
\mathcal{F}_{t};R^{n})$, $\forall \eta \in L^{2}(\mathcal{F}_{s};R^{n})$, $%
\mathcal{\rho }_{s,t}^{g}[\xi +\eta ]=\mathcal{\rho }_{s,t}^{g}[\xi ]-\eta
,\forall s\in \lbrack 0,t]$ iff $g$ does not depend on $y$. The cash
additivity property is also called translation invariance. While the
convexity and the monotonicity axioms have been largely accepted by
academics and practitioners, the cash additive axiom has been criticized
from an economic viewpoint. A basic reason is that while regulators and
financial institutions determine and collect today the reserve amounts to
cover future risky positions, the cash additivity requires that risky
positions and reserve amounts are expressed in the same num$\acute{e}$raire.
This is a stringent requirement that limits the applicability of cash
additive risk measures. Moreover, the cash additive axiom does not allow to
account for stochastic discount factor. \textrm{\cite{er} }relaxed the cash
additivity axiom to cash subadditivity. By Theorem 5.4, a multidimensional $%
g $-risk measure satisfying the cash additive axiom and the monotonicity
axiom is in fact $n$ one dimensional $g$-risk measures put together. For a $%
g $-risk measure with generator $g:=-r_{t}y+$ $\overline{g}(t,z)$, where $%
\overline{g}(t,0)=0$, the corresponding $g$-risk measure $\rho
_{t,T}^{g}[\cdot ]$ satisfies the following translation invariance axiom: $%
\forall t\in \lbrack 0,T]$, $\forall C\in R^{n},\xi \in L^{2}(\mathcal{F}%
_{T};R^{n})$,%
\begin{equation}
\mathcal{\rho }_{t,T}^{g}[\xi +C\exp (\int_{0}^{T}r_{s}ds)]=\mathcal{\rho }%
_{t,T}^{g}[\xi ]-C\exp (\int_{0}^{t}r_{s}ds).
\end{equation}

\subsection{Model uncertainty and robust measurement of risks}

When making a decision, different traders may use different probability
measures. When measuring the risk of a financial position, we may face
several models to choose. In these cases, we are interested in knowing the
minimum and maximum of a family of risk measures.

Assume that there is a family of multidimensional $g$-risk measures $\rho
_{s,t}^{g^{\alpha }}$ with parameters ($g^{\alpha },\xi ^{\alpha }$)
satisfying (H1) $\thicksim $(H3) and condition \eqref{3.4}. We will show
that, under some mild conditions, the supremum $\mathrm{ess\ }\underset{%
\alpha }{\sup }\rho _{s,t}^{g^{\alpha }}\left[ \xi ^{\alpha }\right] $ is
also a $g$-risk measure with parameters ($\mathrm{ess\ }\underset{\alpha }{%
\sup }g^{\alpha }$, $\mathrm{ess\ }\underset{\alpha }{\sup }\xi ^{\alpha }$).

\begin{theorem}
Suppose that there is an $\overline{\alpha }$ such that $\forall (y,z)$,
P-a.s., t-a.e.%
\begin{equation}
g(t,y,z)=\mathrm{ess\ }\underset{\alpha }{\sup }g^{\alpha }(t,y,z)=g^{%
\overline{\alpha }}(t,y,z),
\end{equation}%
\begin{equation}
\xi =\mathrm{ess\ }\underset{\alpha }{\sup }\xi ^{\alpha }=\xi ^{\overline{%
\alpha }}.
\end{equation}%
Let $\rho _{s,t}^{g}\left[ \xi \right] $ be the solution of BSDE ($g,\xi $).
Then%
\begin{equation*}
\rho _{s,t}^{g}\left[ \xi \right] =\mathrm{ess\ }\underset{\alpha }{\sup }%
\rho _{s,t}^{g^{\alpha }}\left[ \xi ^{\alpha }\right] =\rho _{s,t}^{g^{%
\overline{\alpha }}}\left[ \xi ^{\overline{\alpha }}\right] ,\forall t\in
\lbrack 0,T],P\text{-}a.s..
\end{equation*}
\end{theorem}

\textbf{Proof}.$\ $ Since ($g^{\alpha },\xi ^{\alpha }$) and ($g,\xi $)
satisfy the requirement of Comparison Theorem, for any $\alpha $, we have $%
\rho _{s,t}^{g}\left[ \xi \right] \geq \rho _{s,t}^{g^{\alpha }}\left[ \xi
^{\alpha }\right] $ and hence $\rho _{s,t}^{g}\left[ \xi \right] \geq
\mathrm{ess\ }\underset{\alpha }{\sup }\rho _{s,t}^{g^{\alpha }}\left[ \xi
^{\alpha }\right] $, $\forall t\in \lbrack 0,T],P$-$a.s.$.

On the other hand, since ($g,\xi $) is attainable at some $\overline{\alpha }
$, $\rho _{s,t}^{g}\left[ \xi \right] $ and $\rho _{s,t}^{g^{\overline{%
\alpha }}}\left[ \xi ^{\overline{\alpha }}\right] $ are both solutions of
BSDE ($g^{\overline{\alpha }},\xi ^{\overline{\alpha }}$). By the Uniqueness
Theorem we get that%
\begin{equation*}
\mathrm{ess\ }\underset{\alpha }{\sup }\rho _{s,t}^{g^{\alpha }}\left[ \xi
^{\alpha }\right] \leq \rho _{s,t}^{g}\left[ \xi \right] =\rho _{s,t}^{g^{%
\overline{\alpha }}}\left[ \xi ^{\overline{\alpha }}\right] \leq \mathrm{%
ess\ }\underset{\alpha }{\sup }\rho _{s,t}^{g^{\alpha }}\left[ \xi ^{\alpha }%
\right] .
\end{equation*}%
Hence we obtain the desired result. $\square $

This theorem still holds when \textquotedblleft $\sup $\textquotedblright\
is replaced by \textquotedblleft $\inf $\textquotedblright . It is a
generalization of dual representation for convex and coherent $g$-risk
measures in which $\left\{ {g^{\alpha }}\right\} $ are linear generators.
Thus we conclude that

$\bullet $ Convex and coherent $g$-risk measures are robust risk measures;

$\bullet $ Model uncertainty could lead to the convexity of risk measures.

Of course, model uncertainty we consider here is under one same
probabilistic framework (or different probability measures which are
equivalent to the reference probability measure). \textrm{\cite{p4} }%
considered coherent risk measures with different probabilistic models which
are mutually singular in the language of $G$-expectation. See also \textrm{%
\cite{bk} }for convex and non-dominated risk measures.

\subsection{Motivations and Applications}

Multidimensional BSDEs have been found several practical applications, among
others, see mean-variance hedging in \textrm{\cite{mt}}, nonzero game in
\textrm{\cite{eh},} switching problem in \textrm{\cite{ht}}, \textrm{\cite%
{hz}}\ and references therein. In this section we shall show how our
multidimensional approach can be applied to measure the insolvency risk of
interacted business units.

\subsubsection{Measuring the insolvency risk of a firm with interacted
subsidiaries}

One way to measure the insolvency risk of a firm is to use the probability
of insolvency itself. It is to determine the minimal amount of capital to
add to the firm such that the probability of insolvency is less than or
equal to some given probability $\beta $. That is to find the minimal $%
\alpha $ such that $P\left\{ X+\alpha (1+r)<0\right\} \leq \beta $, where X
is the risky asset hold in time $T$ for the firm, $r$ is the interest rate
between $0$ and $T$. The solution to the above problem is called Value at
Risk (VaR). It is shown that VaR is not a time consistent risk measure by
\textrm{\cite{adeh2}. }They studied various common risk measures including
VaR. However they omitted an intuitive risk measure that one could
construct: the cost of an insurance policy protecting the firm against
insolvency. \textrm{\cite{Jarrow} }defined the premium of a put option on
the firm as a measure of insolvency risk. We recall Jarrow's idea
detailedly. A firm is in fact a portfolio of assets and liabilities. The
firm's net value is the value of the firm's assets minus liabilities. A
measure of insolvency risk is the cost of buying a put option written on the
firm' net value with zero strike price. Such a put option could guarantee
that the firm's net value will always be greater than or equal to zero.

Let us be more precise. Let $(\Omega ,\mathcal{F},P)$ be a probability
space. Let $\mathcal{G}_{T}$ denote the space of all financial positions
with maturity $T$, i.e., the set of $R^{n}$-valued functions defined on $%
\Omega $. Let $r$ denote the interest rate between time $0$ and $T$. A firm
is defined by its balance sheet at time $T$. The balance sheet consists of
assets $\mathcal{A}\in \mathcal{G}_{T}$, liabilities $\mathcal{L\in G}_{T}$,
and equity $\mathcal{E\in G}_{T}$ such that the following accounting
identities hold:%
\begin{equation}
\mathcal{E}=\mathcal{A}-\mathcal{L}\text{, if }\mathcal{A}-\mathcal{L}\geq 0%
\text{ and }\mathcal{E}=0\text{ if }\mathcal{A}-\mathcal{L}<0.
\end{equation}%
We call $\widetilde{X}=\mathcal{A}-\mathcal{L}$ the firm's net value. Note
that it maybe negative while $\mathcal{E}$ cannot due to its limited
liability status. We say the firm is insolvent if $\widetilde{X}$ is
strictly negative, or else the firm is solvent. Capital relates to the
amount of the firm's net value invested in the riskless asset (e.g. cash).
It is a different notion from equity. Let X be the risky assets minus
liabilities and $\alpha $ be the firm's capital invested at time 0. Then the
firm's net value $\widetilde{X}=X+\alpha (1+r)$. We want to quantify the
insolvency risk of the firm's net value $\widetilde{X}=X+\alpha (1+r)$. As
illustrated above, VaR is a such measure. Now we consider a put option
written on the firm's net value $\widetilde{X}$ with strike price zero. The
terminal condition is
\begin{equation}
\max \left[ -\left( X+\alpha (1+r)\right) ,0\right] .
\end{equation}%
Under the condition of arbitrage-free, \textrm{\cite{Jarrow} }defined the
\textquotedblleft Put Premium Risk Measure\textquotedblright\ as
\begin{equation}
\rho \lbrack \widetilde{X}]=\frac{E\left( \max \left[ -\widetilde{X},0\right]
\right) }{1+r}.  \label{5.24}
\end{equation}%
It measures the economic cost of avoiding insolvency. This is a single
period (static) model. It can be generalized to the dynamic situation.

Recall that the classical Black-Scholes formula corresponds to the following
BSDE:%
\begin{equation}
dY_{t}=r_{t}Y_{t}+\theta _{t}Z_{t}dt+Z_{t}dB_{t},\ Y_{T}=\xi ,\ t\in \lbrack
0,T],  \label{5.25}
\end{equation}%
where $\left( r_{t}\right) $ is the instantaneous interest rate, $\left(
\theta _{t}\right) $ is called a risk premium and ($Z_{t}$) is viewed as the
portfolio of risk investment. The meaning of \eqref{5.25} is that the change
of the value of an option equals to the interest income plus the return of
risk investment and plus a random noise. See \textrm{\cite{epq} }for a
derivation. We then can define a dynamic $g$-risk measure $\rho
_{t,T}^{g}[\xi ]$ with $\xi =\max \left[ -\left( X+\alpha (1+r)\right) ,0%
\right] $ and $g\left( t,y,z\right) =-\left( r_{t}y+\theta _{t}z\right) $.
Generally in a financial market (may or may not be complete), we can define
a more general $g$-risk measure as the solution of the following BSDE:%
\begin{equation}
dY_{t}=-g\left( t,Y_{t},Z_{t}\right) dt+Z_{t}dB_{t},\ Y_{T}=\max \left[
-\left( X+\alpha (1+r)\right) ,0\right] ,\ t\in \lbrack 0,T],
\end{equation}%
The above put option provides insurance against the firm's insolvency. When
the firm is insolvent, it pays off and covers the loss $X+\alpha (1+r)<0$.
See \textrm{\cite{epq} }for an abundance of examples corresponding to
various $g$.

The above put premium risk measure can be summarized to more involved cases.
Economic agents such as consumers, producers, banks, and stockholders,
constantly interact with each other in different ways and for different
purposes. They may compete, imitate and communicate with each other or
suspect strategies of their counterparties and somehow out of these
individual interactions a certain coherence at the aggregate level develops (%
\textrm{\cite{km}}). The first contribution in which the problem of
stochastic interaction was explicitly treated is that of \textrm{\cite{fm}}.
He showed that if the characteristics of agents, for example their
preferences are random but dependent on those of others, the effect of large
numbers of agents is not enough to eliminate uncertainty at the aggregate
level.

Indeed economic agents are continuously learning and changing their
strategies. One is interested in knowing how to measure risks they
encounter, in particular the insolvency risk for a firm with a number of
interacted subsidiaries. See the following examples.

\begin{example}
Consider a firm with two subsidiaries competing against each other for
finite resource $(L_{t})$ (e.g. cash income of the firm) at each time $t$ to
hedge their possible net values $\xi ^{1}$ and $\xi ^{2}$ at time $T$
respectively. Let ($r_{t}^{1},r_{t}^{2}$) and ($\theta _{t}^{1},\theta
_{t}^{2}$) be their risk exposure to the spot interest and risky investment.
Let ($Y_{t}^{1},Y_{t}^{2}$) and ($Z_{t}^{1},Z_{t}^{2}$) be their hedging
processes and portfolios of risk investment. The first agent suspects his
counterparty may get $\alpha _{t}^{2}Y_{t}^{2}$ from the finite resource $%
(L_{t})$ with proportional coefficient $0\leq \alpha _{t}^{2}\leq 1$. Thus
he gets $(L_{t}-\alpha _{t}^{2}Y_{t}^{2})^{+}$. And the second agent
suspects his counterparty, the first agent may get $\alpha _{t}^{1}Y_{t}^{1}$
from the finite resource $(L_{t})$ with proportional coefficient $0\leq
\alpha _{t}^{1}\leq 1$. Thus he gets $(L_{t}-\alpha _{t}^{1}Y_{t}^{1})^{+}$.
To reduce their insolvency risks, the two subsidiaries can decide whether or
not to buy a put option with payoff $\max \left[ -\xi ^{i},0\right] $, i=1,2
at time T. Then a natural problem comes up: how to estimate this put option?
Using Jarrow's formula \eqref{5.24} or the classical Black-Scholes equation %
\eqref{5.25}? Obviously different pricing formulas or pricing under various
markets (complete or incomplete) yield different current values. The current
value is the capital a subsidiary needs to protect herself against
insolvency. An appropriate and efficient way is that the two subsidiaries
estimate the put options by themselves according their own flows of cash
payments\thinspace /\thinspace incomes. That is, they spend the required
capitals (equal to prices of the put options) on investment such that the
risk capitals hedge the insolvency risks at time T. Therefore in a complete
financial market the pricing processes of the two put options should satisfy%
\begin{equation}
dY_{t}^{1}=\left( r_{t}^{1}Y_{t}^{1}+(L_{t}-\alpha
_{t}^{2}Y_{t}^{2})^{+}+\theta _{t}^{1}Z_{t}^{1}\right)
dt+Z_{t}^{1}dB_{t},Y_{T}^{1}=\max \left[ -\xi ^{1},0\right] ,\ t\in \lbrack
0,T],
\end{equation}%
\begin{equation}
dY_{t}^{2}=\left( r_{t}^{2}Y_{t}^{2}+(L_{t}-\alpha
_{t}^{1}Y_{t}^{1})^{+}+\theta _{t}^{2}Z_{t}^{2}\right)
dt+Z_{t}^{2}dB_{t},Y_{T}^{2}=\max \left[ -\xi ^{2},0\right] ,\ t\in \lbrack
0,T].
\end{equation}%
Then ($\rho _{t}^{1}[\xi ^{1}],\rho _{t}^{2}[\xi ^{2}]$):=($%
Y_{t}^{1},Y_{t}^{2}$) is a 2-dimensional risk measure but not sublinear. $%
Y_{t}^{1}$ and $Y_{t}^{2}$ are interconnected in their generators because of
their reciprocity of benefits. The firm should allocate corresponding
capitals (equal to $\rho _{t}^{1}$, $\rho _{t}^{2}$ resp.) at time t to her
subsidiaries to protect them against insolvency.
\end{example}

See also the following simple example.

\begin{example}
Consider a multi-business firm with two business units holding stocks of
each other for sharing their Profit$\&$Loss. The first unit holds $20\%$
stocks of the second one and the second unit holds $10\%$ stocks of the
first one. Thus their insolvency risks can be measured by%
\begin{equation*}
dY_{t}^{1}=\left[ r_{t}\left( 0.9Y_{t}^{1}+0.2Y_{t}^{2}\right) +\theta
_{t}Z_{t}^{1}\right] dt+Z_{t}^{1}dB_{t},\ Y_{T}^{1}=\max \left[ -\xi ^{1},0%
\right] ,
\end{equation*}%
\begin{equation*}
dY_{t}^{2}=\left[ r_{t}\left( 0.1Y_{t}^{1}+0.8Y_{t}^{2}\right) +\theta
_{t}Z_{t}^{2}\right] dt+Z_{t}^{2}dB_{t},\ Y_{T}^{2}=\max \left[ -\xi ^{2},0%
\right] .
\end{equation*}
\end{example}

It is interesting that \textrm{\cite{st}} extended various risk measures in
discrete time to the solutions of BSDEs, including VaR, semi-deviation risk
measure, Average VaR and Gini risk measure. Thus, BSDEs provide an abundance
of dynamic valuations (resp. risk measure) in continuous time. For instance,
the entropic risk measure is the most famous example of a convex risk
measure, introduced by \textrm{\cite{fs2}}. \textrm{\cite{be}} pointed out
that the classical dynamic entropic risk measure under a Brownian filtration
corresponds to a quadratic BSDE with generator $g=\frac{1}{2\gamma }%
\left\vert z\right\vert ^{2}$, where $\gamma >0$ is the risk tolerance
coefficient.

Along with the rapid development of the financial market, risk relationships
will be more and more complicated. For instance, a set of related options, a
group of players in a game, companies in the same industry, due to countless
ties among them, it is more applicable to consider multidimensional risk
measures to control their risks.

\subsubsection{Acceptance sets and insurance $g$-risk measures}

From an insolvency point of view, an acceptance set identifies those risk
positions that are acceptable such that the firm is solvent. Any risk
measure $\rho :\mathcal{G_{T}\mapsto }R^{n}$ induces an acceptance set
\begin{equation*}
A_{\rho }:=\left\{ \xi \in \mathcal{G}_{T}|\ \rho \lbrack \xi ]\leq
0\right\} .
\end{equation*}%
Conversely, for a given acceptance set $\mathcal{A}$, the associated risk
measure can be defined as%
\begin{equation*}
\rho \lbrack \xi ]:=\inf \left\{ \alpha \in R^{\mathit{n}}|\ \xi +\alpha \in
A\right\} .
\end{equation*}%
Within the static framework of risk measures, \textrm{\cite{adeh2} }%
presented four axioms for an acceptance set should satisfy. And they
summarized the relations between a coherent risk measure and its acceptance
set. In this section we put emphasis on the following definition:%
\begin{equation}
\widetilde{\rho }_{t,T}^{g}[\xi ]:=\mathrm{ess}\inf \left\{ \eta \in L^{2}(%
\mathcal{F}_{t};R^{\mathit{n}})|\ \rho _{t,T}^{g}[\xi +\eta ]\leq 0\right\}
,\ \xi \in L^{2}(\mathcal{F}_{T};R^{\mathit{n}}),  \label{5.29}
\end{equation}%
where $\rho _{t,T}^{g}$ is a $g$-risk measure. The measure $\widetilde{\rho }%
_{t,T}^{g}$ is the minimal capital required to protect the firm against
insolvency at time $t$. We call it \textquotedblleft Insurance Risk
Measure\textquotedblright\ as in \textrm{\cite{Jarrow}. }Of course if we
define the following acceptance set%
\begin{equation*}
A_{\rho ^{g}}^{t}:=\left\{ \xi \in L^{2}(\mathcal{F}_{T};R^{\mathit{n}%
})|\rho _{t,T}^{g}[\xi ]\leq 0\right\} ,
\end{equation*}%
then $\widetilde{\rho }_{t,T}^{g}$ can be also written as%
\begin{equation*}
\widetilde{\rho }_{t,T}^{g}[\xi ]:=\mathrm{ess}\inf \left\{ \eta \in L^{2}(%
\mathcal{F}_{t};R^{\mathit{n}})|\ \xi +\eta \in A_{\rho ^{g}}^{t}\right\} ,\
\xi \in L^{2}(\mathcal{F}_{T};R^{\mathit{n}}).
\end{equation*}%
If $\rho _{t,T}^{g}$ is convex in the sense of Definition 5.3, it is easy to
prove that $A_{\rho ^{g}}^{t}$ is convex, i.e. $\forall \xi ^{1},\xi ^{2}\in
A_{\rho ^{g}}^{t}$, $\forall \lambda \in \lbrack 0,1]$, we have $\left(
\lambda \xi ^{1}+(1-\lambda )\xi ^{2}\right) \in A_{\rho ^{g}}^{t}$. We are
interested in the relation between $\rho _{t,T}^{g}$ and $\widetilde{\rho }%
_{t,T}^{g}$. Obviously we have

\begin{proposition}
Let $\rho _{t,T}^{g}$ be a multidimensional dynamic $g$-risk measure. Then

(i) $\widetilde{\rho }_{t,T}^{g}$ is always translation invariant.

(ii) $\widetilde{\rho }_{t,T}^{g}\equiv \rho _{t,T}^{g}$ if and only if $g$
does not depend on y.

(iii) $\widetilde{\rho }_{t,T}^{g}\geq \rho _{t,T}^{g}$ if and only if $\rho
_{t,T}^{g}$ satisfies the axiom of cash subadditivity:
\begin{equation}
\rho _{t,T}^{g}[\xi +\eta ]\geq \rho _{t,T}^{g}[\xi ]-\eta ,\forall \eta \in
L^{2}(\mathcal{F}_{t};R^{\mathit{n}})\mathbf{,}\forall \xi \in L^{2}(%
\mathcal{F}_{T};R^{\mathit{n}}).  \label{5.30}
\end{equation}
\end{proposition}

The static risk measure $\widetilde{\rho }_{0,T}^{g}$ induced by $\rho
_{0,T}^{g}$ performs even \textquotedblleft better\textquotedblright\ than
the original one because it is always translation invariant and
constant-keeping. It extends the notion of \textquotedblleft Insurance Risk
Measure\textquotedblright\ in \textrm{\cite{Jarrow}} to generally nonlinear
case.

It seems to be more reasonable to establish the axiom of cash subadditivity
as%
\begin{equation}
\rho _{t,T}^{g}[\xi +\eta ]\geq \rho _{t,T}^{g}[\xi ]-\eta ,\forall \eta \in
L^{2}(\mathcal{F}_{t};R_{+}^{\mathit{n}})\mathbf{,}\forall \xi \in L^{2}(%
\mathcal{F}_{T};R^{\mathit{n}}).  \label{5.31}
\end{equation}%
An example in the following (Example 5.3) shows that the typical
risk-neutral BSDE verifies \eqref{5.31} but not \eqref{5.30}. See \textrm{%
\cite{be} }for other motivating examples. Now we consider the following
(nonnegative) insurance $g$-risk measure:%
\begin{equation}
\widetilde{\rho }_{t,T}^{+g}[\xi ]:=\mathrm{ess}\inf \left\{ \eta \in L^{2}(%
\mathcal{F}_{t};R_{+}^{\mathit{n}})|\ \rho _{t,T}^{g}[\xi +\eta ]\leq
0\right\} ,\ \xi \in L^{2}(\mathcal{F}_{T};R^{\mathit{n}}),  \label{5.32}
\end{equation}

\begin{theorem}
Let $g$ satisfy (H1) $\thicksim $(H3) and \eqref{3.4} and $g(\cdot ,\mathbf{0%
},\mathbf{0})=\mathbf{0}$.\ Assume also that $\rho _{t,T}^{g}$ satisfies the
cash subadditivity \eqref{5.31}. Then (i) $\widetilde{\rho }_{t,T}^{+g}[\xi
]=0$, $\forall \xi \geq 0$; (ii) $\widetilde{\rho }_{t,T}^{+g}[\cdot ]$ is
nonincreasing; (iii) $\widetilde{\rho }_{t,T}^{+g}[\xi ]\geq \rho
_{t,T}^{g}[\xi ]$,$\ \forall \xi \in L^{2}(\mathcal{F}_{T};R^{\mathit{n}})$;
(iv) $\widetilde{\rho }_{t,T}^{+g}[\xi +\eta ]\geq \widetilde{\rho }%
_{t,T}^{+g}[\xi ]-\eta ,\forall \eta \in L^{2}(\mathcal{F}_{t};R_{+}^{%
\mathit{n}})\mathbf{,}\forall \xi \in L^{2}(\mathcal{F}_{T};R^{\mathit{n}}).$
\end{theorem}

The proofs are trivial. It is interesting that $\widetilde{\rho }_{t,T}^{+g}$
may be not time consistent.

\begin{example}
Consider the 1-dimensional risk-neutral BSDE ($\xi ,g,T$) with $g\left(
y\right) =-ry$, $r\in R^{\mathit{+}}$, $y\in R$. The induced g-risk measure $%
\rho _{t,T}^{g}[\xi ]=e^{-r\left( T-t\right) }\cdot E[-\xi |\mathcal{F}_{t}]$
satisfies the cash subadditivity \eqref{5.31}. It is easy to deduce that $%
\widetilde{\rho }_{t,T}^{+,g}[\xi ]=\left( E[-\xi |\mathcal{F}_{t}]\right)
^{+}$ and it satisfies Theorem 5.6. For $\xi =-B_{T}$ and $t\in (0,T]$, we
have
\begin{equation*}
\widetilde{\rho }_{0,t}^{+,g}\left[ -\widetilde{\rho }_{t,T}^{+,g}[-B_{T}]%
\right] =\widetilde{\rho }_{0,t}^{+,g}\left[ -B_{t}^{+}\right]
=E[B_{t}^{+}]>0=\left( E[B_{T}]\right) ^{+}=\widetilde{\rho }_{0,T}^{+,g}%
\left[ -B_{T}\right] .
\end{equation*}%
Thus $\widetilde{\rho }_{t,T}^{+,g}$ violates the axiom of time consistency
for some random variables.
\end{example}

Risk management is different from the firm's and the regulator's
perspective. From the point of view of supervision, regulators prefer $%
\widetilde{\rho }_{t,T}^{+g}$ to $\rho _{t,T}^{g}$ while the firms may just
choose $\rho _{t,T}^{g}$ to prepare their risk capital.

\subsubsection{Optimal risk sharing for $\protect\gamma $-tolerant $g$-risk
measures}

The topic of optimal risk sharing has been studied early by \textrm{\cite%
{borch}} in the insurance and reinsurance context and recently by \textrm{%
\cite{be05,be}} for cash additive $\gamma $-tolerant convex risk measures.
For non-homogeneous convex risk measures, the impact of the size of the
position is not linear. It seems natural to consider the relation between
risk tolerance and the perception of the size of positions. The risk
tolerance coefficient $\gamma $ describes how agent penalize compared with
the root risk measure $\rho $. Precisely a $\gamma $-tolerant risk measure
is defined as%
\begin{equation*}
\rho _{t,T}^{\gamma }\left[ \xi \right] :=\gamma \rho _{t,T}[\frac{\xi }{%
\gamma }],\gamma \in R^{+}.
\end{equation*}%
A typical example is the entropic risk measure $e_{t,T}^{\gamma }\left[ \xi %
\right] =\gamma \ln {\normalsize E}[\exp (-\frac{\xi }{\gamma })|\mathcal{F}%
_{t}]$.

The optimal risk sharing between two economic parties, denoted by A and B
means that: A want to issue a financial product $\eta \in L^{2}(\mathcal{F}%
_{T})$ and sell it to B for price $\pi _{t}$ at time $t$. A tries to
minimize her risk
\begin{equation*}
\mathrm{ess\ }\underset{\eta ,\pi }{\inf }\left\{ \rho _{t,T}^{A}[\xi -\eta
]-\pi _{t}\right\} \text{ subject to }\rho _{t,T}^{B}[\eta ]+\pi _{t}\leq 0%
\text{.}
\end{equation*}%
The price of $\eta $ is determined by B as $\pi _{t}=-\rho _{t,T}^{B}[\eta ]$%
. Hence B should have an interest in doing the transaction. At least the
product $\eta $ does not worsen her risk. Then the minimization problem
becomes%
\begin{equation*}
\mathrm{ess\ }\underset{\eta }{\inf }\left\{ \rho _{t,T}^{A}[\xi -\eta
]+\rho _{t,T}^{B}[\eta ]\right\} \text{,}
\end{equation*}%
which is called inf-convolution of two functionals and denoted by $\rho
^{A}\square \rho ^{B}[\xi ]$.

Now we consider two groups A and B, measuring their risks by a
multidimensional $g$-risk measure $\rho ^{g}$ with different risk tolerant
coefficients $\gamma _{A},\gamma _{B}\in R^{+}$, i.e., their risks are
measured by $\rho _{t,T}^{g_{\gamma _{A}}}:=$ $\gamma _{A}\rho _{t,T}^{g}[%
\frac{\cdot }{\gamma _{A}}]$ and $\rho _{t,T}^{g_{\gamma _{B}}}:=$ $\gamma
_{B}\rho _{t,T}^{g}[\frac{\cdot }{\gamma _{B}}]$ respectively.

\begin{theorem}
Let $\rho _{t,T}^{g}$ be a multidimensional convex $g$-risk measure with g
satisfying (H1) $\thicksim $(H3) and \eqref{3.4} and $g(\cdot ,\mathbf{0},%
\mathbf{0})=\mathbf{0}$.\ Then $\forall \gamma _{A},\gamma _{B}\in R^{+}$,

(i) $\rho ^{g_{\gamma _{A}}}\square \rho ^{g_{\gamma _{B}}}$ is a $\left(
\gamma _{A}+\gamma _{B}\right) $-tolerant $g$-risk measure, i.e., it is the
unique solution of the following BSDE%
\begin{equation}
Y_{t}=-\xi +\int_{t}^{T}\left( \gamma _{A}+\gamma _{B}\right) g\left( s,%
\frac{Y_{s}}{\left( \gamma _{A}+\gamma _{B}\right) },\frac{Z_{s}}{\left(
\gamma _{A}+\gamma _{B}\right) }\right) ds-\int_{t}^{T}Z_{s}dB_{s},\ 0\leq
t\leq T.  \label{5.320}
\end{equation}

(ii) $\left( \rho ^{g_{\gamma }}\right) _{\gamma =\infty }$ is a $g_{\infty
} $-risk measure with generator $g_{\infty }=\underset{\gamma \uparrow
+\infty }{\lim }\gamma g\left( t,\frac{y}{\gamma },\frac{z}{\gamma }\right) $%
. And for different $g,\overline{g}$ satisfying the above assumptions, we
have%
\begin{equation*}
\rho ^{g_{\infty }}\square \rho ^{\overline{g}_{\infty }}=\left( \rho
^{g}\square \rho ^{\overline{g}}\right) _{\infty }.
\end{equation*}
\end{theorem}

\textbf{Proof}. (i) By the convexity of $\rho _{t,T}^{g}$ and $\rho
_{t,T}^{g}[\mathbf{0}]=\mathbf{0}$, it is easy to check that $\rho
_{t,T}^{g_{\gamma }}[\xi ]$ is nonincreasing in $\gamma $ and $%
p_{t,T}^{g}\left( \gamma ,\xi \right) :=\rho _{t,T}^{g_{\gamma }}[\xi ]$ is
sublinear (positive-homogeneous and subadditive) in $\left( \gamma ,\xi
\right) $. Thus for any $\eta \in L^{2}(\mathcal{F}_{T};R^{\mathit{n}})$,
\begin{equation}
\left( \gamma _{A}+\gamma _{B}\right) \rho _{t,T}^{g}[\frac{\xi }{\left(
\gamma _{A}+\gamma _{B}\right) }]\leq \gamma _{A}\rho _{t,T}^{g}[\frac{\xi
-\eta }{\gamma _{A}}]+\gamma _{B}\rho _{t,T}^{g}[\frac{\eta }{\gamma _{B}}],
\label{5.321}
\end{equation}%
and consequently $\left( \gamma _{A}+\gamma _{B}\right) \rho _{t,T}^{g}[%
\frac{\xi }{\left( \gamma _{A}+\gamma _{B}\right) }]\leq \mathrm{ess\ }%
\underset{\eta }{\inf }\left\{ \gamma _{A}\rho _{t,T}^{g}[\frac{\xi -\eta }{%
\gamma _{A}}]+\gamma _{B}\rho _{t,T}^{g}[\frac{\eta }{\gamma _{B}}]\right\} $%
. When taking $\eta =\frac{\gamma _{B}\xi }{\left( \gamma _{A}+\gamma
_{B}\right) }$, inequality \eqref{5.321} becomes an equality. Hence $\left(
\gamma _{A}+\gamma _{B}\right) \rho ^{g}[\frac{\xi }{\left( \gamma
_{A}+\gamma _{B}\right) }]=\rho ^{g_{\gamma _{A}}}\square \rho ^{g_{\gamma
_{B}}}$. Obviously $\left( \gamma _{A}+\gamma _{B}\right) \rho _{t,T}^{g}[%
\frac{\xi }{\left( \gamma _{A}+\gamma _{B}\right) }]$ is the unique solution
of BSDE \eqref{5.320}. Furthermore, $\eta ^{\ast }=\frac{\gamma _{B}\xi }{%
\left( \gamma _{A}+\gamma _{B}\right) }$ is the optimal transfer.

(ii) Since $g$ is Lipschitz (thus dominated by positive-homogeneous
functions, see Remark 3.1) and both $\rho ^{g_{\gamma }}$ and $g_{\gamma
}:=\gamma g\left( t,\frac{y}{\gamma },\frac{z}{\gamma }\right) $ are
nonincreasing in $\gamma $, $g_{\infty }$ and $\rho ^{g_{\infty }}$ are well
defined. Then by Theorem 5.5, $\rho ^{g_{\infty }}$ is the solution of BSDE
with generator $g_{\infty }$. Since $\rho ^{g_{\gamma }}$ and $\rho ^{%
\overline{g}_{\gamma }}$ are nonincreasing in $\gamma $, by a direct
calculation, we have%
\begin{eqnarray*}
\rho ^{g_{\infty }}\square \rho ^{\overline{g}_{\infty }} &=&\mathrm{ess\ }%
\underset{\eta }{\inf }\ \mathrm{ess\ }\underset{\gamma }{\inf }\{\gamma
\rho _{t,T}^{g}[\frac{\xi -\eta }{\gamma }]+\gamma \rho _{t,T}^{\overline{g}%
}[\frac{\eta }{\gamma }]\} \\
&=&\mathrm{ess\ }\underset{\gamma }{\inf }\ \mathrm{ess\ }\underset{\eta }{%
\inf }\{\gamma (\rho _{t,T}^{g}[\frac{\xi }{\gamma }-\eta ]+\rho _{t,T}^{%
\overline{g}}[\eta ])\} \\
&=&\mathrm{ess\ }\underset{\gamma }{\inf }\rho ^{g_{\gamma }}\square \rho ^{%
\overline{g}_{\gamma }}=\left( \rho ^{g}\square \rho ^{\overline{g}}\right)
_{\infty }.
\end{eqnarray*}%
$\square $

The financial meaning of (i) is obvious: the behavior of risk sharing
increases A's risk tolerance capacity which equals to the sum of the risk
tolerance of both groups.

\subsection{Other ways to obtain $g$-risk measures}

There is another way to define risk measure via $g$-expectation.

\begin{definition}
Suppose that the risk mechanism $g$ satisfies (H1) $\thicksim $(H3) and %
\eqref{3.4} and $g(\cdot ,\mathbf{0},\mathbf{0})=\mathbf{0}$.\ For any $t\in
\lbrack 0,T]$,$\ $the risk position $\xi \in L^{2}(\mathcal{F}_{t};R^{n})$,\
we define $\overline{\mathcal{\rho }}^{g}[\xi ]=-\mathcal{E}_{g}^{t}[\xi ]$,
$\overline{\mathcal{\rho }}_{s,t}^{g}[\xi ]=-\mathcal{E}_{g}^{t}[\xi |%
\mathcal{F}_{s}]$, $s\in \lbrack 0,t]$
\end{definition}

Obviously $\overline{\mathcal{\rho }}^{g}[\xi ]\ $and $\overline{\mathcal{%
\rho }}_{s,t}^{g}[\xi ]$ are multidimensional static risk measure and
multidimensional dynamic risk measure respectively. However they have
different meanings compared to $g$-risk measures in Definition 5.2. For a
future risk position $\xi $ with maturity time $t$, if we give the price of $%
\xi $ via $g$-expectation and view $\mathcal{E}_{g}^{t}[\xi ]$ as the
estimation value, then the risk measure $\overline{\mathcal{\rho }}^{g}[\xi
] $ means to measure the risk of $\xi $ via the opposite value of its
estimation $\mathcal{E}_{g}^{t}[\xi ]$. A natural question comes up: for a
specific risk position $\xi $, we use which mechanism to measure its risk,
or equivalently which mechanism is more robust.

\subsubsection{Multidimensional $g$-pricing mechanisms}

In an option market, the ask-bid pricing mechanism is operated through the
system of market makers. The ask price and the bid price often do not
coincide with each other (see \textrm{\cite{jn00}, \cite{bn09b}}). If $%
\mathcal{E}_{g}^{T}[\xi |\mathcal{F}_{t}]$ is the ask price at time $t$ of a
derivative $\xi $ with maturity $T$, then the bid price must be $-\mathcal{E}%
_{g}^{T}[-\xi |\mathcal{F}_{t}]$ and we have generally $\mathcal{E}%
_{g}^{T}[\xi |\mathcal{F}_{t}]\geq -\mathcal{E}_{g}^{T}[-\xi |\mathcal{F}%
_{t}]$. A $g$-expectation is called a $g$-pricing mechanism here. We have

\begin{theorem}
Let (H1) $\thicksim $(H3) hold for $g$. For any $t\in \lbrack 0,T]$, $%
\forall \xi \in L^{2}(\mathcal{F}_{t};R^{n})$,\ $\mathcal{E}_{g}^{t}[\xi |%
\mathcal{F}_{s}]\geq -\mathcal{E}_{g}^{t}[-\xi |\mathcal{F}_{s}]$, $\forall
s\in \lbrack 0,t]$, if and only if for any $k=1,2,...,n$, $\forall t\in
\lbrack 0,T]$, $\forall y^{\prime }\in R^{n}$,
\begin{equation}
g_{k}(t,\delta ^{k}y+y^{\prime },z)\geq -g_{k}(t,-y^{\prime },-z^{\prime }),
\label{5.33}
\end{equation}%
for any $\delta ^{k}y\in R^{n}$ such that $\delta ^{k}y\geq 0$, $(\delta
^{k}y)_{k}=0$, and $z,z^{\prime }\in R^{n\times d}$ with $(z)_{k}=(z^{\prime
})_{k}$.
\end{theorem}

\textbf{Proof}.$\ $ Let ($\mathcal{E}_{g}^{t}[-\xi |\mathcal{F}_{s}]$)
denote the solution of the following BSDE:

\begin{equation}
\overline{Y}_{s}=-\xi +\int_{s}^{t}g\left( r,\overline{Y}_{r},\overline{Z}%
_{r}\right) dr-\int_{s}^{t}\overline{Z}_{r}dB_{r},\ 0\leq s\leq t.
\label{5.34}
\end{equation}%
Then $(-\mathcal{E}_{g}^{t}[-\xi |\mathcal{F}_{s}])$ solves the following
BSDE:
\begin{equation}
\widetilde{Y}_{s}=\xi +\int_{s}^{t}-g\left( r,-\widetilde{Y}_{r},-\widetilde{%
Z}_{r}\right) dr-\int_{s}^{t}\widetilde{Z}_{r}dB_{r},\ 0\leq s\leq t.
\label{5.35}
\end{equation}%
Applying Theorem 3.1 to the above BSDE \eqref{5.35} and BSDE \eqref{2.2}, it
is deduced that for all derivative $\xi $, $\mathcal{E}_{g}^{t}[\xi |%
\mathcal{F}_{s}]\geq -\mathcal{E}_{g}^{t}[-\xi |\mathcal{F}_{s}]$, $\forall
s\in \lbrack 0,t]$ if and only if for any $k=1,2,...,n$, $\forall t\in
\lbrack 0,T]$, $\forall y^{\prime }\in R^{n}$, $g_{k}(t,\delta
^{k}y+y^{\prime },z)\geq -g_{k}(t,-y^{\prime },-z^{\prime })$, for any $%
\delta ^{k}y\in R^{n}$ such that $\delta ^{k}y\geq 0$, $(\delta ^{k}y)_{k}=0$%
, and $z,z^{\prime }\in R^{n\times d}$ with $(z)_{k}=(z^{\prime })_{k}$. $%
\square $

An axiomatic approach in continuous time asset pricing is developed in
\textrm{\cite{jn00}. }The axiomatization of dynamic pricing procedure in the
setting of Brownian filtration is presented in \textrm{\cite{p2,p3}} in the
language of\textrm{\ }$g$-expectation. Pricing via dynamic convex risk
measures has also been studied by \textrm{\cite{jr}. \cite{bn09b} }%
introduced a notion of time consistent dynamic pricing procedure on a
filtered probability space which assigns to every essentially bounded
financial position a dynamic ask price and a bid price, taking transaction
costs and liquidity risk into account. See also \textrm{\cite{bn08b}}.%
\textrm{\ }Characterization of one dimensional $g$-pricing mechanisms by its
generating function was studied sufficiently in \textrm{\cite{p3}}. For a
risk mechanism $g$, if inequality \eqref{5.33} holds, then we have $\rho
_{t}^{g}[\xi ]\geq \overline{\mathcal{\rho }}_{t}^{g}[\xi ]$, $\forall \xi
\in L^{2}(\mathcal{F}_{T};R^{\mathit{n}})$. Thus usually we utilize $\rho
_{t}^{g}[\xi ]$ rather than $\overline{\mathcal{\rho }}_{t}^{g}[\xi ]$ to
measure the risk of a financial position $\xi $ in an option market.

\subsubsection{Utility-based $g$-risk measure}

Let us consider a small agent who can consume between time $0$ and time $T$.
Let $c_{t}$ be the (positive) consumption rate at time $t$. We assume that
there exists a terminal reward $\xi $ at time $T$. The utility at time $t$
is a function of the instantaneous consumption rate $c_{t}$ and of the
future utility (corresponding to the future consumption). In fact, the
recursive utility is assumed to satisfy the following BSDE in form of
conditional expectation (\textrm{\cite{de1, de2}}),
\begin{equation*}
u_{t}=E[\xi +\int_{t}^{T}g\left( s,c_{s},u_{s}\right) ds|\mathcal{F}_{t}].
\end{equation*}%
We consider a more general class of recursive utilities defined by Peng's
BSDE (\textrm{\cite{epq}}),
\begin{equation}
u_{t}=\xi +\int_{t}^{T}g\left( s,c_{s},u_{s}^{1}\ldots
u_{s}^{n},z_{s}\right) ds-\int_{t}^{r}z_{s}dB_{s}.  \label{5.36}
\end{equation}

In general, utilities must satisfy the following classical properties: (a)
Monotonicity with respect to the terminal value and to the consumption. (b)
Concavity with respect to the consumption. (c) Time consistency: this means
that, for any two consumption processes $c^{1}$and $c^{2}$ and any time $t$,
if $c^{1}$\ and $c^{2}$ are identical up to time $t$ and if the continuation
of $c^{1}$ is preferred to the continuation of $c^{2}$ at time $t$, then $%
c^{1}$ is preferred to $c^{2}$ at time $0$. Certainly by Theorem 3.1 and
Theorem 5.2, if $g$ is concave w.r.t $(c,y,z)$, quasi-monotonously
increasing w.r.t $(c,y)$, and for any $(c,y)$, the generator $g_{k}$ does
not depend on $\left( z_{j}\right) _{j\neq k}$, then the above properties
are satisfied. From a financial point of view, the generator $g\left(
t,u_{t},z_{t}\right) $ represents the instantaneous utility at time $t$ of
consumption rate $c_{t}$. The dependence of $g_{k}$ on $u^{1},\cdots ,u^{n}$
can be interpreted as a nonzero-sum game problem, where the players'
utilities affect each other and consequently the generators are
interconnected.

Let $u_{t}$ be a function of utility, a utility-based risk measure is
defined by $\rho _{t}=-u_{t}$ (see \textrm{\cite{ku}} for example). Hence we
define the utility-based $g$-risk measure $\overline{\mathcal{\rho }}%
_{t}^{g}[\cdot ]$ by the solution $u_{t}^{g}$ of BSDE \eqref{5.36}. Note
that unlike $\rho _{t}^{g}[\cdot ]$, the convexity (subadditivity) of risk
measure $\overline{\mathcal{\rho }}_{t}^{g}[\cdot ]$ is not consistent with
its generator $g$, i.e.: if $g$ is concave (superadditive), then $\overline{%
\mathcal{\rho }}_{t}^{g}[\cdot ]$ is convex (resp. subadditive). Thus for
investigating convex utility-based risk measure, one usually studies concave
utility (\textrm{\cite{ku}}).

Observe that by Theorem 5.4, when $g$ does not depend on $y$, the definition
$\overline{\mathcal{\rho }}_{t,T}^{g}[\xi ]=-\mathcal{E}_{g}^{T}[\xi |%
\mathcal{F}_{s}]$ coincides with the following
\begin{equation}
\hat{\rho}_{t,T}^{g}[\xi ]:=\mathrm{ess}\inf \left\{ \eta \in L^{2}(\mathcal{%
F}_{t};R^{\mathit{n}})|\mathcal{E}_{g}^{T}[\xi +\eta |\mathcal{F}_{t}]\geq
0\right\} .  \label{5.37}
\end{equation}%
So generally a risk measure based on a recursive utility is not consistent
with definition \eqref{5.37}. Based on some experimental evidence, \textrm{%
\cite{bw}} argued that in general a risk measure should not be considered,
up to a sign, an utility functional. Let $u$ be a concave, strictly
increasing utility function with $u(0)=0$. The risk measure $\rho ^{u}$ is
defined as $\rho ^{u}[\xi ]:=\inf \left\{ \alpha \in R|\ E_{P}\left[ u\left(
\xi +\alpha \right) \right] \geq 0\right\} $. \textrm{\cite{fsd} }also gave
a partial support which shows that the preference structure defined by $%
E_{P}u$ does not coincide with the reverse ordering provided by $\rho ^{u}$,
unless $u$ is linear or exponential. Time consistency is a crucial property
for utility-based risk measures. \textrm{\cite{ck09}} gave some equivalent
conditions for time consistency of a dynamic utility function. It is shown
that the agent's indifference prices are recursive if and only if the
preferences are translation invariant.

\begin{remark}
Four definitions of g-risk measures are referred in this paper. They are%
\begin{equation*}
\rho _{t,T}^{g}[\xi ]:=\mathcal{E}_{g}^{T}[-\xi |\mathcal{F}_{t}],\ \xi \in
L^{2}(\mathcal{F}_{T};R^{\mathit{n}});
\end{equation*}%
\begin{equation*}
\overline{\mathcal{\rho }}_{t,T}^{g}[\xi ]:=-\mathcal{E}_{g}^{T}[\xi |%
\mathcal{F}_{t}],\ \xi \in L^{2}(\mathcal{F}_{T};R^{\mathit{n}});
\end{equation*}%
\begin{equation*}
\hat{\rho}_{t,T}^{g}[\xi ]:=\mathrm{ess}\inf \left\{ \eta \in L^{2}(\mathcal{%
F}_{t};R^{\mathit{n}})|\ \mathcal{E}_{g}^{T}[\xi +\eta |\mathcal{F}_{t}]\geq
0\right\} ,\ \xi \in L^{2}(\mathcal{F}_{T};R^{\mathit{n}});
\end{equation*}%
\begin{equation*}
\widetilde{\rho }_{t,T}^{g}[\xi ]:=\mathrm{ess}\inf \left\{ \eta \in L^{2}(%
\mathcal{F}_{t};R^{\mathit{n}})|\ \mathcal{E}_{g}^{T}[-\left( \xi +\eta
\right) |\mathcal{F}_{t}]\leq 0\right\} ,\ \xi \in L^{2}(\mathcal{F}_{T};R^{%
\mathit{n}}).
\end{equation*}%
If $g(t,y,z)=-g(t,-y,-z)$, then $\rho _{t,T}^{g}=\overline{\mathcal{\rho }}%
_{t,T}^{g}$; if g does not depend on y, then $\overline{\mathcal{\rho }}%
_{t,T}^{g}=\hat{\rho}_{t,T}^{g}$, $\rho _{t,T}^{g}=\widetilde{\rho }%
_{t,T}^{g}$; if g does not depend on y and $g(t,z)=-g(t,-z)$, then four
definitions coincide with each other. Generally they are different
definitions. We will continue our work to give insights among them and seek
new applications in particular for the last two.
\end{remark}

\section{CONCLUSION}

During a long time it seems that many problems related to multidimensional
BSDEs are expected to be complicated because of the lack of a clear
structure of multidimensional BSDEs. This paper, by virtue of certain
mathematical tools, establishes some properties of multidimensional $g$-risk
measures. Particularly we show that a multidimensional $g$-risk measure is
nonincreasingly convex if and only if the generator $g$ satisfies a
quasi-monotone increasingly convex condition.

Indeed $g$-risk measures represent a large class of risk measures. In one
dimensional case, any dynamic risk measure satisfying certain domination
condition can be written as a conditional $g$-expectation (\textrm{\cite%
{chmp}}, \textrm{\cite{p2}}). As to the multidimensional case, similar
result is presented in the appendix. Nevertheless, $g$-risk measures are
interesting by themselves which are strongly linked to a functional $g$.
Choosing carefully the coefficient $g$ enables to generate dynamic $g$-risk
measures that are locally compatible with the views and practice of
different agents in the market. Especially a general dual representation of
a multidimensional dynamic convex $g$-risk measure is obtained in which the
penalty term is expressed explicitly via the polar function of $g$. It is
shown that model uncertainty could lead to the convexity of risk measures.

Our multidimensional approach provides an alternative way to measure the
insolvency risk of a firm with interacted subsidiaries. As pointed out by
\textrm{\cite{Jarrow}}, the cost of buying a put option written on the
firm's net value is an intuitive measure of the insolvency risk of a firm.
Optimal risk sharing between two groups with different risk tolerant
coefficients is also investigated. We hope that this theory will be included
in the toolbox of standard financial software in a near future. Different
ways to obtain $g$-risk measures are compared at the end of the paper.

Finally, there are also several interesting issues such as optimal risk
transfer and risk allocation within a group with interacted subsidiaries, we
do not consider due to the length of the paper. We will solve these issues
in our future publications.

\appendix

\section{BSDE REPRESENTATION FOR MULTIDIMENSIONAL INTERACTED RISK MEASURES
AND NONLINEAR EXPECTATIONS}

This section will show that, in the setting of Brownian filtration, if a
multidimensional interacted risk measure $\mathcal{\rho }_{s,t}$ satisfying
certain domination condition, then it is in fact a $g$-risk measure, i.e. it
can be represented by a multidimensional BSDE. For a multidimensional risk
measure $\mathcal{\rho }_{s,t}=\left( \mathcal{\rho }_{s,t}^{1},\ldots ,%
\mathcal{\rho }_{s,t}^{n}\right) $, we assume its components are interacted
with each other, or else it is just a notion putting $n$ $1$-dimensional
risk measures together. Thus, for any $k=1,\ldots ,n$, the $k$th component $%
\mathcal{\rho }_{s,t}^{k}$ is in fact a functional of the other dimensions,
i.e., $\mathcal{\rho }_{s,t}^{k}$ depends on the whole path of $\left(
\mathcal{\rho }^{l}\right) _{l\neq k}$ on the time interval $[s,t]$.
Precisely, we write $\mathcal{\rho }_{s,t}^{k}\left[ \cdot \right] _{\left(
\mathcal{\rho }^{1},\ldots ,\mathcal{\rho }^{k-1},\mathcal{\rho }%
^{k+1}\ldots ,\mathcal{\rho }^{n}\right) }$ instead of $\mathcal{\rho }%
_{s,t}^{k}\left[ \cdot \right] $ to put emphasis on the dependence of one
dimension on the other dimensions. Sometimes for simplification we write an $%
R^{n-1}$-valued vector $\left( y^{l}\right) _{l\neq k}:=(y^{1},\ldots
,y^{k-1},y^{k+1}\ldots ,y^{n})$. We make the following assumptions:

({\normalsize A6}) $\forall k=1,\ldots ,n$, $\forall \left( y_{.}^{l}\right)
_{l\neq k}\in L_{\mathcal{F}}^{2}(0,T;R^{n-1})$, $\mathcal{\rho }_{s,t}^{k}%
\left[ \cdot \right] _{\left( y_{.}^{l}\right) _{l\neq k}}$ is well defined
on $L^{2}(\mathcal{F}_{t};R)$ and it satisfies (A1) $\thicksim $(A4).

The above condition means that fixing other dimensions, the $k$th dimension
is still a risk measure satisfying (A1) $\thicksim $(A4) except (A5). The
condition of $\mathcal{\rho }^{g_{\mu }}$-domination is

({\normalsize A7}) For any $k=1,\ldots ,n$, $\forall \xi ^{k_{1}},\xi
^{k_{2}}\in L^{2}(\mathcal{F}_{t};R)$, $\forall \left( x_{s}^{1},\ldots
,x_{s}^{n}\right) $, $\left( y_{s}^{1},\ldots ,y_{s}^{n}\right) \in L_{%
\mathcal{F}}^{2}(0,T;R^{n})$,%
\begin{equation*}
\mathcal{\rho }_{s,t}^{k}\left[ \xi ^{k_{1}}\right] _{\left(
x_{.}^{l}\right) _{l\neq k}}-\mathcal{\rho }_{s,t}^{k}\left[ \xi ^{k_{2}}%
\right] _{\left( y_{.}^{l}\right) _{l\neq k}}\leq \mathcal{\rho }%
_{s,t}^{g_{\mu },k}\left[ \xi ^{k_{1}}-\xi ^{k_{2}}\right] _{\left(
x_{.}^{l}-y_{.}^{l}\right) _{l\neq k}},
\end{equation*}%
where $\mathcal{\rho }_{s,t}^{g_{\mu },k}\left[ \xi ^{k}\right] _{\left(
y_{.}^{l}\right) _{l\neq k}}$ is the solution of the following 1-dimensional
BSDE:%
\begin{equation}
Y_{s}^{k}=-\xi ^{k}+\int_{s}^{t}\mu \left( \sum_{l\neq k}\left\vert
y_{r}^{l}\right\vert +\left\vert Y_{r}^{k}\right\vert +\left\vert
Z_{r}^{k}\right\vert \right) dr-\int_{s}^{t}Z_{r}^{k}dB_{r},\ s\in \lbrack
0,t].  \label{5.21}
\end{equation}

\begin{theorem}
Let ({\normalsize A1}) $\thicksim $({\normalsize A7}) hold for the
multidimensional dynamic risk measure $\rho _{s,t}[\cdot ]$. Then there
exists a function $g:\Omega \times \lbrack 0,T]\times R^{n}\times R^{n\times
\mathit{d}}\longmapsto R^{n}$ satisfying the \textquotedblleft
quasi-monotonicity condition\textquotedblright\ \eqref{3.4} and the
Lipschitz condition ({\normalsize H2}) and $g(\cdot ,\mathbf{0},\mathbf{0})=%
\mathbf{0}$, such that for any $k=1,\ldots ,n$, $\forall \xi \in L^{2}(%
\mathcal{F}_{t};R^{n})$,
\begin{equation*}
\rho _{s,t}[\xi ]=\rho _{s,t}^{g}[\xi ].
\end{equation*}
\end{theorem}

We now present the proof in the language of nonlinear expectation.

We define the nonlinear expectation operator $\mathcal{E}_{s.t}\left[ \xi %
\right] :=\rho _{s.t}\left[ -\xi \right] $, which means that $\mathcal{E}%
_{s.t}$ satisfies (\textbf{A'1}) Monotonicity. $\forall t\in \lbrack 0,T]$, $%
\mathcal{E}_{s,t}[\xi ^{1}]\geq \mathcal{E}_{s,t}[\xi ^{2}]$, if $\forall
\xi ^{1}\geq \xi ^{2}\in L^{2}(\mathcal{F}_{t};R^{n})$; (\textbf{A}$\prime $%
\textbf{2}) $\mathcal{E}_{t,t}[\xi ]=\xi $, $\forall \xi \in L^{2}(\mathcal{F%
}_{t};R^{n})$; (\textbf{A'3}) Time-consistency. $\mathcal{E}_{s,T}[\xi ]=%
\mathcal{E}_{s,t}^{g}[\mathcal{E}_{t,T}^{g}[\xi ]]$, $\forall 0\leq s\leq
t\leq T$, $\forall \xi \in L^{2}(\mathcal{F}_{T};R^{n})$; (\textbf{A'4})
Regularity. $\mathbf{1}_{A}\mathcal{E}_{s,t}[\xi \mathbf{1}_{A}]=\mathbf{1}%
_{A}\mathcal{E}_{s,t}[\xi ]$, $\forall A\in \mathcal{F}_{s}$; (\textbf{A'5})
Normalization. $\mathcal{E}_{s,t}[\mathbf{0}]=\mathbf{0}$; (\textbf{A'6}) $%
\forall k=1,\ldots ,n$, for any $\left( n-1\right) $-dimensional vector
process $\left( y_{s}^{1},\ldots ,y_{s}^{k-1},y_{s}^{k+1}\ldots
,y_{s}^{n}\right) \in L_{\mathcal{F}}^{2}(0,T;R^{n})$, $\mathcal{E}_{s,t}^{k}%
\left[ \cdot \right] _{\left( y_{s}^{1},\ldots
,y_{s}^{k-1},y_{s}^{k+1}\ldots ,y_{s}^{n}\right) }$ is well defined on $%
L^{2}(\mathcal{F}_{t};R)$ and it satisfies (A'1) $\thicksim $(A'4); (\textbf{%
A'7}) $\mathcal{E}^{g_{\mu }}$-domination. For any $k=1,\ldots ,n$, $\forall
\xi ^{k_{1}},\xi ^{k_{2}}\in L^{2}(\mathcal{F}_{t};R)$, $\forall \left(
x_{s}^{1},\ldots ,x_{s}^{n}\right) ,\left( y_{s}^{1},\ldots
,y_{s}^{n}\right) \in L_{\mathcal{F}}^{2}(0,T;R^{n})$,%
\begin{equation}
\mathcal{E}_{s,t}^{k}\left[ \xi ^{k_{1}}\right] _{\left( x_{.}^{l}\right)
_{l\neq k}}-\mathcal{E}_{s,t}^{k}\left[ \xi ^{k_{2}}\right] _{\left(
y_{.}^{l}\right) _{l\neq k}}\leq \mathcal{E}_{s,t}^{g_{\mu },k}\left[ \xi
^{k_{1}}-\xi ^{k_{2}}\right] _{\left( x_{.}^{l}-y_{.}^{l}\right) _{l\neq k}}.
\label{A.1}
\end{equation}%
where $\mathcal{E}_{s,t}^{g_{\mu },k}\left[ \xi ^{k}\right] _{\left(
y_{.}^{l}\right) _{l\neq k}}$ is the solution of \eqref{5.21} with terminal
datum $\xi ^{k}\in L^{2}(\mathcal{F}_{t};R)$.

The condition (A'5) means that $\mathcal{E}_{s,t}^{k}\left[ 0\right]
_{\left( 0,\ldots ,0,0\ldots ,0\right) }=0$, which combining \eqref{A.1}
yields that for any $k=1,\ldots ,n$, $\forall \xi ^{k}\in L^{2}(\mathcal{F}%
_{t};R)$, $\forall \left( y_{s}^{1},\ldots ,y_{s}^{n}\right) \in L_{\mathcal{%
F}}^{2}(0,T;R^{n})$,%
\begin{equation}
\mathcal{E}_{s,t}^{k}\left[ \xi ^{k}\right] _{\left( y_{.}^{l}\right)
_{l\neq k}}\leq \mathcal{E}_{s,t}^{g_{\mu },k}\left[ \xi ^{k}\right]
_{\left( y_{.}^{l}\right) _{l\neq k}}.
\end{equation}%
Let $\mathcal{E}_{s,t}^{-g_{\mu },k}\left[ \xi ^{k}\right] _{\left(
y_{.}^{l}\right) _{l\neq k}}$, $k=1,\ldots n$, denotes the solution of the
following $1$-dimensional BSDE:%
\begin{equation}
Y_{s}^{k}=\xi ^{k}-\int_{s}^{t}\mu \left( \sum_{l\neq k}\left\vert
y_{r}^{l}\right\vert +\left\vert Y_{r}^{k}\right\vert +\left\vert
Z_{r}^{k}\right\vert \right) dr-\int_{s}^{t}Z_{r}^{k}dB_{r},\ s\in \lbrack
0,t].
\end{equation}%
Observing that $\mathcal{E}_{s,t}^{-g_{\mu },k}\left[ \cdot \right] _{\left(
y_{.}^{l}\right) _{l\neq k}}=-\mathcal{E}_{s,t}^{g_{\mu },k}\left[ -\cdot \ %
\right] _{\left( y_{.}^{l}\right) _{l\neq k}}$. Then by \eqref{A.1}, one can
directly check that:%
\begin{equation}
\mathcal{E}_{s,t}^{-g_{\mu },k}\left[ \xi ^{k_{1}}-\xi ^{k_{2}}\right]
_{\left( x_{.}^{l}-y_{.}^{l}\right) _{l\neq k}}\leq \mathcal{E}_{s,t}^{k}%
\left[ \xi ^{k_{1}}\right] _{\left( x_{.}^{l}\right) _{l\neq k}}-\mathcal{E}%
_{s,t}^{k}\left[ \xi ^{k_{2}}\right] _{\left( y_{.}^{l}\right) _{l\neq k}}
\end{equation}%
and%
\begin{equation}
\mathcal{E}_{s,t}^{-g_{\mu },k}\left[ \xi ^{k}\right] _{\left(
y_{.}^{l}\right) _{l\neq k}}\leq \mathcal{E}_{s,t}^{k}\left[ \xi ^{k}\right]
_{\left( y_{.}^{l}\right) _{l\neq k}}.
\end{equation}

Let $D_{\mathcal{F}}^{2}(0,T;R^{n})$ denote all the processes in $L_{%
\mathcal{F}}^{2}(0,T;R^{n})$ with right-continuous and with left-limit
(RCLL) paths such that $E\left[ \sup_{0\leq t\leq T}|\varphi _{t}|^{2}\right]
<\infty $.

\begin{definition}
Let $\mathcal{E}_{s.t}$ be an n-dimensional nonlinear expectation. A process
$\left( Y_{s}\right) \in D_{\mathcal{F}}^{2}(0,T;R^{n})$ is called $\mathcal{%
E}$-supermartingale ($\mathcal{E}$-submartingale) if
\begin{equation}
\mathcal{E}_{s,t}^{k}\left[ Y_{t}\right] _{\left( Y_{.}^{l}\right) _{l\neq
k}}\leq Y_{s}^{k}\text{, (\textquotedblleft }\geq \text{\textquotedblright\
resp.\ ), }\forall s\in \left[ 0,t\right] ,\text{k=1,\ldots n.}
\end{equation}
\end{definition}

Obviously if $\mathcal{E}_{s.t}$ is $\mathcal{E}_{s,t}^{g_{\mu }}$%
-dominated, then $\mathcal{E}_{s.t}$ is an $\mathcal{E}_{s,t}^{g_{\mu }}$%
-submartingale and an $\mathcal{E}_{s,t}^{-g_{\mu }}$-supermartingale and $%
\mathcal{E}_{s,t}^{g_{\mu }}$ is an $\mathcal{E}_{s,t}$-supermartingale.

For a multidimensional nonlinear expectation operator $\mathcal{E}$ and a
process $A\in D_{\mathcal{F}}^{2}(0,T;R^{n})$ and a random variable $\xi \in
$ $L^{2}(\mathcal{F}_{T};R^{n})$, we write $\mathcal{E}_{t,T}\left[ \xi ;A%
\right] :=\mathcal{E}_{t,T}\left[ \xi +\left( A_{T}-A_{t}\right) \right] $
for brevity. To prove Theorem A.1, we need the following nonlinear
Doob-Meyer decomposition.

\begin{proposition}
Let (H1) $\thicksim $(H3) hold for $g$ and $g_{k}$ does not depend on $%
\left( z_{j}\right) _{j\neq k}$. Let $\left( Y_{t}\right) \in D_{\mathcal{F}%
}^{2}(0,T;R^{n})$ be an $\mathcal{E}_{t,T}^{g}$-supermartingale. Then there
exists a unique increasing process $\left( A_{s}\right) \in D_{\mathcal{F}%
}^{2}(0,T;R^{n})$ with $A_{0}=0$ such that%
\begin{equation}
Y_{t}=\mathcal{E}_{t,T}^{g}\left[ Y_{T};A\right] ,
\end{equation}%
i.e. for any $k=1,\ldots ,n$,
\begin{equation}
Y_{t}^{k}=Y_{T}^{k}+\int_{t}^{T}g^{k}\left( s,Y_{s},Z_{s}^{k}\right)
ds+\left( A_{T}^{k}-A_{t}^{k}\right) -\int_{t}^{T}Z_{s}^{k}dB_{s},\ 0\leq
t\leq T,  \label{A.8}
\end{equation}
\end{proposition}

\textbf{Proof}.$\ $ The main idea is to construct the following family of
BSDEs parameterized by $m$=1,2,\ldots
\begin{equation}
^{k}y_{t}^{m}=Y_{T}^{k}+\int_{t}^{T}g^{k}\left( s,Y_{s}^{1},\ldots
,Y_{s}^{k-1},{}^{k}y_{s}^{m},Y_{s}^{k+1}\ldots
,Y_{s}^{n},{}{}^{k}z_{s}^{m}\right)
ds+m\int_{t}^{T}(Y_{s}^{k}-{}^{k}y_{s}^{m})ds-%
\int_{t}^{T}{}^{k}z_{s}^{m}dB_{s},  \label{A.9}
\end{equation}%
where$\ k=1,\ldots ,n$, to approximate the solution triple $\left(
Y,Z,A\right) $ of BSDE \eqref{A.8}. Note that \eqref{A.9} is in fact a
1-dimensional BSDE. Then we can show that the sequence $\left\{
^{k}y_{t}^{m}\right\} _{m=1}^{\infty }$ is increasing in $m$ and bounded by $%
^{k}y_{t}^{1}$ and $Y_{t}^{k}$ in $L_{\mathcal{F}}^{2}(0,T;R)$ and $%
^{k}y_{t}^{m}$ converges up to the process $\left( Y_{t}^{k}\right) $. The
proof is an analogy of Lemma 3.4 and Theorem 3.3 in \textrm{\cite{p99}}. We
omit it here for brevity.. $\square $

\begin{corollary}
For a given $\left( C_{t}\right) \in D_{\mathcal{F}}^{2}(0,T;R^{n})$, if $%
\left( Y_{t}\right) \in D_{\mathcal{F}}^{2}(0,T;R^{n})$ is an $\mathcal{E}%
_{t,T}^{g}\left[ \cdot ;C\right] $-supermartingale, i.e.
\begin{equation}
^{k}\mathcal{E}_{s,t}^{g}\left[ Y_{t}^{k};C^{k}\right] _{\left(
Y_{.}^{l}\right) _{l\neq k}}\leq Y_{s}^{k},\text{k=1,\ldots n.}
\end{equation}%
Then there exists a unique increasing process $\left( A_{s}\right) \in D_{%
\mathcal{F}}^{2}(0,T;R^{n})$ with $A_{0}=0$ such that%
\begin{equation}
Y_{t}=\mathcal{E}_{t,T}^{g}\left[ Y_{T};C+A\right] .
\end{equation}
\end{corollary}

\bigskip Since $\left( Y_{s}^{t,X,C}\right) _{s\in \lbrack 0,t]}$ is an $%
\mathcal{E}_{s,t}^{g_{\mu }}\left[ \cdot ;C\right] $-submartingale and an $%
\mathcal{E}_{s,t}^{-g_{\mu }}\left[ \cdot ;C\right] $-supermartingale, as a
consequence of Corollary A.1 and condition \eqref{A.1}, we have

\begin{lemma}
Let (A'1) $\thicksim $(A'7) hold. Given a process $\left( C_{t}\right) \in
D_{\mathcal{F}}^{2}(0,T;R^{n})$, for the process $Y_{s}^{t,X,C}:=\mathcal{E}%
_{s,t}\left[ X;C\right] $, with $s\in \lbrack 0,t]$, $X\in L^{2}(\mathcal{F}%
_{t};R^{n})$, there exist a $\left( g_{s}^{t,X,C}\right) \in L_{\mathcal{F}%
}^{2}(0,t;R^{n})$ and a $\left( Z_{s}^{t,X,C}\right) \in L_{\mathcal{F}%
}^{2}(0,t;R^{n\times d})$ such that, for any $k=1,\ldots ,n$,%
\begin{equation}
^{k}Y_{s}^{t,X,C}=X^{k}+\left( C_{t}^{k}-C_{s}^{k}\right)
+\int_{s}^{t}{}^{k}g_{r}^{t,X,C}dr-\int_{s}^{t}{}^{k}Z_{r}^{t,X,C}dB_{r},\
0\leq s\leq t,  \label{A.14}
\end{equation}%
and
\begin{equation}
\left\vert ^{k}g_{s}^{t,X,C}\right\vert \leq \mu \left( \left\vert
Y_{s}^{t,X,C}\right\vert +\left\vert ^{k}Z_{s}^{t,X,C}\right\vert \right)
,\forall s\in \lbrack 0,t].  \label{A.15}
\end{equation}%
Moreover if ($X,C$) changes, then
\begin{equation}
\left\vert ^{k}g_{s}^{t,X,C}-{}^{k}g_{s}^{t,X^{\prime },C^{\prime
}}\right\vert \leq \mu \left( \left\vert Y_{s}^{t,X,C}-Y_{s}^{t,X^{\prime
},C^{\prime }}\right\vert +\left\vert
^{k}Z_{s}^{t,X,C}-^{k}Z_{s}^{t,X^{\prime },C\prime }\right\vert \right)
,\forall s\in \lbrack 0,t].  \label{A.16}
\end{equation}
\end{lemma}

We refer to Proposition 6.6 in \textrm{\cite{p2}} for similar arguments.%
\textrm{\ }

\begin{corollary}
For $t^{\prime }\in \lbrack t,T]$, $X^{\prime }\in L^{2}(\mathcal{F}%
_{t^{\prime }};R^{n})$, $\left( C_{t}^{\prime }\right) \in D_{\mathcal{F}%
}^{2}(0,T;R^{n})$, let $Y_{s}^{t^{\prime },X^{\prime },C^{\prime }}:=%
\mathcal{E}_{s,t^{\prime }}\left[ X^{\prime };C^{\prime }\right] $. Then%
\begin{equation*}
\left\vert ^{k}g_{s}^{t,X,C}-^{k}g_{s}^{t^{\prime },X^{\prime },C^{\prime
}}\right\vert \leq \mu \left( \left\vert Y_{s}^{t,X,C}-Y_{s}^{t^{\prime
},X^{\prime },C^{\prime }}\right\vert +\left\vert
^{k}Z_{s}^{t,X,C}-{}^{k}Z_{s}^{t^{\prime },X^{\prime },C\prime }\right\vert
\right) ,\forall s\in \lbrack 0,t].
\end{equation*}
\end{corollary}

\textbf{Proof}.$\ $It is just a consequence of the time consistency: $%
Y_{s}^{t^{\prime },X^{\prime },C^{\prime }}=\mathcal{E}_{s,t}\left[
Y_{t}^{t^{\prime },X^{\prime },C^{\prime }};C^{\prime }\right] ,s\in \lbrack
0,t]$ and Lemma A.1. $\square $

We now present a general nonlinear Doob-Meyer decomposition for $\mathcal{E}$%
-supermartingale.

\begin{proposition}
Let (A'1) $\thicksim $(A'7) hold. Then an $\mathcal{E}$-supermartingale $%
\left( Y_{t}\right) \in D_{\mathcal{F}}^{2}(0,T;R^{n})$ has a unique
decomposition
\begin{equation}
Y_{t}=\mathcal{E}_{t,T}\left[ Y_{T};A\right] ,t\in \left[ 0,T\right] ,
\end{equation}%
where $\left( A_{s}\right) \in D_{\mathcal{F}}^{2}(0,T;R^{n})$ is an
increasing process with $A_{0}=0$.
\end{proposition}

The proof of the above proposition is quite similar to Proposition A.1 and
to Theorem 8.1 in \textrm{\cite{p2}}. We need to construct the following
sequence parameterized by $m$=1,2,\ldots
\begin{equation}
^{k}y_{t}^{m}=\mathcal{E}_{t,T}\left[ Y_{T}^{k}+m%
\int_{t}^{T}(Y_{s}^{k}-{}^{k}y_{s}^{m})ds\right] _{\left( Y_{s}^{l}\right)
_{l\neq k}},\ k=1,\ldots ,n.
\end{equation}%
The extension of $\mathcal{E}_{t,T}$ to $\mathcal{E}_{\sigma ,\tau \text{ }}$%
under stopping times is necessary. Then following the procedure in Theorem
8.1 in \textrm{\cite{p2}}, the proof could be complete.

\textbf{Proof of Theorem A.1}: For fixed $\left( t,y,z\right) \in \lbrack
0,T]\times R^{n}\times R^{n\times d}$, consider the following n-dimensional
SDE:%
\begin{equation}
d^{k}Y_{s}^{t,y,z_{k}}=-{}\mu (\sum_{l=1}^{n}\left\vert
^{l}Y_{s}^{t,y,z_{l}}\right\vert +\left\vert z_{k}\right\vert
)ds+{}z_{k}dB_{s},s\in \lbrack t,T],  \label{A.17}
\end{equation}%
\begin{equation}
Y_{t}^{t,y,z_{k}}=y_{k},  \label{A.18}
\end{equation}%
where $k=1,\ldots ,n$, $\left( y_{k},z_{k}\right) \in R\times R^{d}$. By the
classical theory of SDE, we have the following estimate:%
\begin{equation}
E[\sum_{k=1}^{n}\left\vert ^{k}Y_{s}^{t,y,z_{k}}-y_{k}\right\vert ^{2}]\leq
c\sum_{k=1}^{n}\left( \left\vert y_{k}\right\vert ^{2}+\left\vert
z_{k}\right\vert ^{2}+1\right) \left( s-t\right) ,\forall s\in \lbrack t,T],
\label{A.19}
\end{equation}%
where $c$ is a universal constant changing from line to line in this section.

By (A'7), $\left( Y_{s}^{t,y,z_{k}}\right) _{s\in \lbrack 0,t]}$ is an $%
\mathcal{E}_{s,t}\left[ \cdot \right] $-supermartingale. Then due to
Proposition A.2, $Y_{s}^{t,y,z_{k}}$ has the following decomposition: for
any $k=1,\ldots ,n$,%
\begin{equation}
^{k}Y_{s}^{t,y,z_{k}}=\mathcal{E}_{s,T}\left[
^{k}Y_{T}^{t,y,z_{k}};A^{t,y,z_{k}}\right] ,s\in \lbrack 0,T],  \label{A.20}
\end{equation}%
where $A^{t,y,z_{k}}\in D_{\mathcal{F}}^{2}(0,T;R^{n})$ is an increasing
process with $A_{0}^{t,y,z_{k}}=0$. Applying Lemma A.1 to \eqref{A.20}, then
there exists a pair ($^{k}g_{s}^{t,y,z_{k}},{}^{k}Z_{s}^{t,y,z_{k}}$)$\in L_{%
\mathcal{F}}^{2}(0,T;R)\times L_{\mathcal{F}}^{2}(0,T;R^{d})$ such that%
\begin{equation}
d^{k}Y_{s}^{t,y,z_{k}}=-{}^{k}g_{s}^{t,y,z_{k}}ds-{}d{}A_{s}^{t,y,z_{k}}+{}{}^{k}Z_{s}^{t,y,z_{k}}dB_{s},s\in \lbrack t,T],
\label{A.21}
\end{equation}%
and for each ($y,z_{k}$), ($y^{\prime },z_{k}^{\prime }$)$\in R^{n}\times
R^{d}$, $t\leq t^{\prime }\in \lbrack 0,T]$,%
\begin{equation}
\left\vert ^{k}g_{s}^{t,y,z_{k}}-{}^{k}g_{s}^{t^{\prime },y^{\prime
},z_{k}^{\prime }}\right\vert \leq \mu (\sum_{l=1}^{n}\left\vert
^{l}Y_{s}^{t,y,z_{l}}-{}^{l}Y_{s}^{t^{\prime },y^{\prime },z_{l}^{\prime
}}\right\vert +\left\vert ^{k}Z_{s}^{t,y,z_{k}}-{}^{k}Z_{s}^{t^{\prime
},y^{\prime },z_{k}^{\prime }}\right\vert ),\forall s\in \lbrack t^{\prime
},T],  \label{A.22}
\end{equation}

\begin{equation}
\left\vert ^{k}g_{s}^{t,y,z_{k}}\right\vert \leq \mu
(\sum_{l=1}^{n}\left\vert ^{l}Y_{s}^{t,y,z_{k}}\right\vert +\left\vert
^{l}Z_{s}^{t,y,z_{k}}\right\vert ),\forall s\in \lbrack t,T].  \label{A.23}
\end{equation}%
For $X\in L^{2}(\mathcal{F}_{t^{\prime }};R^{n})$, let $\overline{Y}%
_{s}^{t^{\prime },X}:=\mathcal{E}_{s,t^{\prime }}\left[ X\right] =\mathcal{E}%
_{s,t^{\prime }}\left[ X;0\right] $. Then by Lemma A.1 and Corollary A.2
again, $\overline{Y}_{s}^{t^{\prime },X}$ has the following decomposition,
for any $k=1,\ldots ,n$,%
\begin{equation}
d^{k}\overline{Y}_{s}^{t^{\prime },X}=-{}^{k}\overline{g}_{s}^{t^{\prime
},X}ds+{}^{k}\overline{Z}_{s}^{t^{\prime },X}dB_{s},\,^{k}\overline{Y}%
_{t^{\prime }}^{t^{\prime },X}=X^{k}\ ,\forall s\in \lbrack 0,t^{\prime }]
\label{A.24}
\end{equation}%
where the pair ($^{k}\overline{g}_{s}^{t^{\prime },X},{}^{k}Z_{s}^{t^{\prime
},X}$)$\in L_{\mathcal{F}}^{2}(0,t^{\prime };R)\times L_{\mathcal{F}%
}^{2}(0,t^{\prime };R^{d})$ and
\begin{equation}
\left\vert ^{k}g_{s}^{t,y,z_{k}}-{}^{k}\overline{g}_{s}^{t^{\prime
},X}\right\vert \leq \mu (\sum_{l=1}^{n}\left\vert
^{l}Y_{s}^{t,y,z_{l}}-{}^{l}\overline{Y}_{s}^{t^{\prime },X}\right\vert
+\left\vert ^{k}Z_{s}^{t,y,z_{k}}-^{k}\overline{Z}_{s}^{t^{\prime
},X}\right\vert ),\forall s\in \lbrack t,t^{\prime }],  \label{A.25}
\end{equation}

Comparing \eqref{A.17} with \eqref{A.21}, we obtain $%
^{k}Z_{s}^{t,y,z_{k}}=z_{k}\mathbf{1}_{[t,T]}\left( s\right) $. Therefore %
\eqref{A.22}, \eqref{A.23} and \eqref{A.25} could be rewritten as%
\begin{equation}
\left\vert ^{k}g_{s}^{t,y,z_{k}}-{}^{k}g_{s}^{t^{\prime },y^{\prime
},z_{k}^{\prime }}\right\vert \leq \mu (\sum_{l=1}^{n}\left\vert
^{l}Y_{s}^{t,y,z_{l}}-{}^{l}Y_{s}^{t^{\prime },y^{\prime },z_{l}^{\prime
}}\right\vert +\left\vert z_{k}-{}z_{k}^{\prime }\right\vert ),\forall s\in
\lbrack t^{\prime },T],  \label{A.26}
\end{equation}

\begin{equation}
\left\vert ^{k}g_{s}^{t,y,z_{k}}\right\vert \leq \mu
(\sum_{l=1}^{n}\left\vert ^{l}Y_{s}^{t,y,z_{l}}\right\vert +\left\vert
z_{k}\right\vert ),\forall s\in \lbrack t,T],  \label{A.27}
\end{equation}%
\begin{equation}
\left\vert ^{k}g_{s}^{t,y,z_{k}}-{}^{k}\overline{g}_{s}^{t^{\prime
},X}\right\vert \leq \mu (\sum_{l=1}^{n}\left\vert
^{l}Y_{s}^{t,y,z_{l}}-{}^{l}Y_{s}^{t^{\prime },X}\right\vert +\left\vert
z_{k}-{}^{k}Z_{s}^{t^{\prime },X}\right\vert ),\forall s\in \lbrack
t,t^{\prime }].  \label{A.28}
\end{equation}%
For each natural number $m\geq 1$, we set $t_{i}^{m}=i2^{-m}T$, $%
i=0,1,2,\ldots 2^{m}$, and define
\begin{equation}
^{k}g^{m}\left( s,y,z_{k}\right)
:=\sum_{i=0}^{2^{m}-1}{}^{k}g_{s}^{t_{i}^{m},y,z_{k}}\cdot \mathbf{1}%
_{[t_{i}^{m},t_{i+1}^{m}]}\left( s\right) ,\ s\in \lbrack 0,T].  \label{A.29}
\end{equation}%
Then $^{k}g^{m}\left( s,y,z_{k}\right) $, $k=1,\ldots ,n$, is $\mathcal{F}%
_{s}$-adapted and

\begin{lemma}
For any $k=1,\ldots ,n$, for each fixed $\left( y,z_{k}\right) \in
R^{n}\times R^{d}$, the sequence $\left\{ ^{k}g^{m}\left( \cdot
,y,z_{k}\right) \right\} _{m=1}^{\infty }$ converges to some $\mathcal{F}%
_{s} $-adapted process $^{k}g\left( \cdot ,y,z_{k}\right) \in L_{\mathcal{F}%
}^{2}(0,T;R)$.
\end{lemma}

\textbf{Proof}.$\ $For two natural numbers $0<m<m^{\prime }$, for each $s\in
\lbrack 0,T]$, there are some integers $i\leq 2^{m}-1$ and $j\leq
2^{m^{\prime }}-1$ such that $s\in \lbrack t_{i}^{m},t_{i+1}^{m}]\cap
\lbrack t_{j}^{m^{\prime }},t_{j+1}^{m^{\prime }}]$. Then by \eqref{A.26}
and \eqref{A.19} we can deduce that%
\begin{equation*}
\underset{s\in \lbrack 0,T]}{\sup }E\left\vert ^{k}g^{m}\left(
s,y,z_{k}\right) -{}^{k}g^{m^{\prime }}\left( s,y,z_{k}\right) \right\vert
^{2}\leq 2nc\mu ^{2}\sum_{k=1}^{n}\left( \left\vert y_{k}\right\vert
^{2}+\left\vert z_{k}\right\vert ^{2}+1\right) \left( 2^{-m}+2^{-m^{\prime
}}\right) T\rightarrow 0,
\end{equation*}%
as $m,m\prime \rightarrow \infty $, which means that $^{k}g^{m}\left(
s,y,z_{k}\right) $ converges to some $\mathcal{F}_{s}$-adapted process $%
^{k}g\left( s,y,z_{k}\right) \in L_{\mathcal{F}}^{2}(0,T;R)$.

\begin{lemma}
For any $k=1,\ldots ,n$, the limit function $^{k}g\left( t,y,z_{k}\right) $
satisfies (i) $\left\vert ^{k}g\left( s,y,z_{k}\right) -{}^{k}g\left(
s,y^{\prime },z_{k}^{\prime }\right) \right\vert \leq \mu (\left\vert
y-y^{\prime }\right\vert +\left\vert z_{k}-z_{k}^{\prime }\right\vert )$, $%
\forall y,y^{\prime }\in R^{n}$, $z_{k},z_{k}^{\prime }\in R^{d}$; (ii) $%
^{k}g\left( s,y=0,z_{k}=0\right) =0$; (iii) $\left\vert ^{k}g\left(
s,y,z_{k}\right) -{}^{k}\overline{g}_{s}^{t,X}\right\vert \leq \mu
(\left\vert y-\overline{Y}_{s}^{t,X}\right\vert +\left\vert z_{k}-{}^{k}%
\overline{Z}_{s}^{t,X}\right\vert )$, $\forall s\in \lbrack 0,t]$.
\end{lemma}

\textbf{Proof}.$\ $When $s\in \lbrack t_{j}^{m},t_{j+1}^{m}]$, by %
\eqref{A.29} and \eqref{A.26} we have%
\begin{equation*}
\left\vert ^{k}g^{m}\left( s,y,z_{k}\right) -{}^{k}g^{m}\left( s,y^{\prime
},z_{k}^{\prime }\right) \right\vert \leq \mu \left( \left\vert y-y^{\prime
}\right\vert +\left\vert z_{k}-z_{k}^{\prime }\right\vert \right) +\mu
(\sum_{k=1}^{n}\left\vert ^{k}Y_{s}^{t_{i}^{m},y,z_{k}}-y_{k}\right\vert
)+\mu (\sum_{k=1}^{n}\left\vert ^{k}Y_{s}^{t_{j}^{m},y^{\prime
},z_{k}^{\prime }}-y_{k}^{\prime }\right\vert ).
\end{equation*}%
Then by \eqref{A.19} we can show that the last two items of the right hand
converge to zero as $m\rightarrow \infty $. Thus (i) is obtained. Combining %
\eqref{A.19} and \eqref{A.27}, we could deduce (ii). Applying \eqref{A.28}
and \eqref{A.19} to $\left\vert ^{k}g\left( s,y,z_{k}\right) -{}^{k}%
\overline{g}_{s}^{t,X}\right\vert $, then (iii) is obtained by passing to
the limit. See an earlier version of this paper in \textit{arxiv} for
complete proofs of the above two lemmas.

\textbf{Continuation of the proof of Theorem }A.1: For each $X\in L^{2}(%
\mathcal{F}_{t};R^{n})$, by \eqref{A.24}, $\overline{Y}_{s}^{t,X}:=\mathcal{E%
}_{s,t}\left[ X\right] $ has the following decomposition:%
\begin{equation*}
^{k}\overline{Y}_{s}^{t,X}=X^{k}+\int_{s}^{t}{}^{k}\overline{g}%
_{r}^{t,X}dr+\int_{s}^{t}{}^{k}\overline{Z}_{r}^{t,X}dB_{r},\ s\in \lbrack
0,t],k=1,\ldots ,n.
\end{equation*}%
By Lemma A.3, the following n-dimensional BSDE,%
\begin{equation*}
^{k}Y_{s}^{t,X}=X^{k}+\int_{s}^{t}{}^{k}g\left(
r,Y_{r}^{t,X},{}^{k}Z_{r}^{t,X}\right)
dr-\int_{s}^{t}{}^{k}Z_{r}^{t,X}dB_{r},\ s\in \lbrack 0,t],k=1,\ldots ,n,
\end{equation*}%
has a unique pair of solution $\left( Y_{s}^{t,X},Z_{s}^{t,X}\right) =\left(
{}^{1}Y_{s}^{t,X},\ldots ,{}^{n}Y_{s}^{t,X},{}^{1}Z_{s}^{t,X},\ldots
,{}^{n}Z_{s}^{t,X}\right) $. Then applying It\^{o}'s formula to $\left\vert
\overline{Y}_{s}^{t,X}-Y_{s}^{t,X}\right\vert ^{2}$, we get%
\begin{equation*}
E\left\vert \overline{Y}_{s}^{t,X}-Y_{s}^{t,X}\right\vert
^{2}+E\int_{s}^{t}\left\vert \overline{Z}_{r}^{t,X}-Z_{r}^{t,X}\right\vert
^{2}dr\leq 2\left( \mu n+\mu ^{2}\right) \int_{s}^{t}E\left\vert \overline{Y}%
_{r}^{t,X}-Y_{r}^{t,X}\right\vert ^{2}dr+\frac{1}{2}E\int_{s}^{t}\left\vert
\overline{Z}_{r}^{t,X}-Z_{r}^{t,X}\right\vert ^{2}dr.
\end{equation*}%
It is a consequence of application of Gronwall's inequality that $\overline{Y%
}_{s}^{t,X}\equiv Y_{s}^{t,X}$. Thus $\mathcal{E}_{s,t}\left[ X\right] =%
\overline{Y}_{s}^{t,X}\equiv Y_{s}^{t,X}=\mathcal{E}_{s,t}^{g}\left[ X\right]
$, where the function $g$ satisfies (i) and (ii) in Lemma A.3. Since $\rho
_{s.t}\left[ -\cdot \ \right] :=\mathcal{E}_{s.t}\left[ \cdot \right] $ is a
quasi-monotone risk measure, inequality \eqref{3.4} holds also for $g$. $%
\square $

Note that Theorem A.1 could not be obtained directly from the one
dimensional case.

\end{document}